\documentclass[prx,aps,twocolumn,showpacs,preprintnumbers,amsmath,amssymb]{revtex4-1}

\usepackage{graphicx}
\usepackage{dcolumn}
\usepackage{hyperref}
\usepackage{amsmath,amssymb}
\usepackage{lineno}

\begin{document}

%useful definitions here

%\preprint{APS/123-QED}
\title{Suppressing The Ferroelectric Switching Barrier in Hybrid Improper Ferroelectrics}
\author{Shutong \surname{Li}}
\author{Turan \surname{Birol}}
\email{tbirol@umn.edu}
\affiliation{Department of Chemical Engineering and Materials Science, University of Minnesota, Minneapolis, Minnesota 55455, USA}
\date{\today}

\begin{abstract}
	Integration of ferroelectric materials into novel technological applications requires low coercive field materials, and consequently, design strategies to reduce the ferroelectric switching barriers. In this first principles study, we show that biaxial strain, which has a strong effect on the ferroelectric ground states, can also be used to tune the switching barrier of hybrid improper ferroelectric Ruddlesden-Popper oxides. We identify the region of the strain -- tolerance factor phase diagram where this intrinsic barrier is suppressed, and show that it can be explained in relation to strain induced phase transitions to nonpolar phases. 
\end{abstract}

\maketitle

%=======================================================================
%%
%%
\section{Introduction}

Since the discovery of ferroelectricity in BaTiO$_3$, perovskite oxides have been heavily studied and utilized in applications as ferroelectric materials. Versatility of the perovskite structure allows a large number of complex oxides to be synthesized, but among those, only a small fraction are ferroelectrics\cite{Benedek2013}. A major breakthrough in perovskite-related ferroelectrics is the discovery of hybrid improper ferroelectricity (HIF) as a materials design route in 2011, which led to an explosion in the predictions of novel ferroelectric oxides \cite{Benedek2011}. Among those, the list of examples that are experimentally verified includes A\textsubscript{3}B\textsubscript{2}O\textsubscript{7} HIFs (Ca,Sr)\textsubscript{3}Ti\textsubscript{2}O\textsubscript{7} \cite{Oh2015}, (Sr,Ca)\textsubscript{3}Sn\textsubscript{2}O\textsubscript{7}, Sr\textsubscript{3}Zr\textsubscript{2}O\textsubscript{7} \cite{Wang2017,Yoshida2018,Yoshida2018-2}, as well as a weak ferromagnetic (Ca\textsubscript{0.69}Sr\textsubscript{0.46}Tb\textsubscript{1.85}Fe\textsubscript{2}O\textsubscript{7}) \cite{Pitcher2015}.

Despite the prediction of ferroelectricity and observation of a polar crystal structure in many compounds, experimentally observing the switching of polarization is challenging. For example, the original HIF Ca$_3$Ti$_2$O$_7$ was reported to have a polar structure 20 years before the idea of HIFs was introduced \cite{Elcombe1991}, but the direct evidence of polarization switching was not observed until 2015 \cite{Oh2015}. The reason behind the absence of switching in these materials was initially believed to be large intrinsic coercive fields, or defects in the materials, which typically increase the coercive field \cite{Mulder2013, Lines2001Book}. The high experimental coercive field is not surprising, because the energy scale that needs to be overcome for switching is considered to be determined by the octahedral rotations, which often have an energy scale significantly higher than that of the ferroelectric distortions in typical perovskite oxides. 
Switching was observed in other HIF materials with coercive fields ranging from 120 to 200 kV$\cdot$cm$^{-1}$ \cite{Wang2017, Oh2015,Yoshida2018-2}, and very recently, the smallest coercive field of 39 kV$\cdot$cm$^{-1}$ was observed in single crystals of Sr$_3$Sn$_2$O$_7$ \cite{Xu2020}. 
Though these coercive fields are comparable to values suitable for integration to silicon chips ($E_c \approx 50$~kV$\cdot$cm$^{-1}$), applications such as high-power actuators and low-voltage logic and memory elements ask for ferroelectrics with robust polarizations that can be switched by a lower coercive field \cite{Scott2007,Cole2014, Xu2018, Liu2018}. 
Ultra-low coercive fields as low as 5~kV$\cdot$cm$^{-1}$ were observed in pulsed laser deposition grown Ca$_3$Ti$_2$O$_7$ thin films, but the reason behind this reduction (and whether it is an intrinsic or an extrinsic effect) is not clarified yet \cite{Li2017}. 

Understanding the intrinsic mechanisms that affect the coercive field of HIF materials, and finding new design strategies to reduce these fields are important for their applications. 
In this paper, we illustrate that strain can be an effective means to achieve this. 
Epitaxial strain, obtained by growing thin films on lattice mismatched substrates, has been used extensively as a way to tune the ferroelectric and dielectric properties of perovskites \cite{Wang2018,Chaturvedi2020}. Both the octahedral rotations, and the proper ferroelectric order parameter are strongly coupled with the biaxial strain in most materials, and strain is shown to change the switching energy barrier of ferroelectrics as well.\cite{Clima2014}
HIFs are shown to undergo interesting structural phase transitions under strain as well\cite{Lu2016}, but there is no detailed study of the switching behavior of HIFs under biaxial strain. The original study on HIFs \cite{Benedek2011} showed that the lowest energy switching path and energy (which is correlated with the coercive field) is strain dependent, but the recent work that illustrate the richness of possible switching paths makes it necessary to re-evaluate the polarization switching behavior of strained HIFs \cite{Nowadnick2016, Munro2018}. 

In this study, we perform density functional theory (DFT) calculations on 13 different A$_3$B$_2$O$_7$ Ruddlesden-Popper compounds to map out the strain-tolerance factor phase diagram, and show that the strain induced non-polar or anti-polar phases emerge in compounds with a finite range of tolerance factors. We then show, by performing nudged elastic band (NEB) calculations, that the intrinsic coherent polarization switching energy barrier decreases as the compounds get closer to phase boundaries by biaxial strain. 
This suppression of switching barrier is not always accompanied with a decrease in the polarization, which makes strain tuning of HIF Ruddlesden-Poppers a viable tool to obtain low coercive field ferroelectrics with a robust polarization. We also show that the tensile and compressive strains favor different switching pathways, which can be intuitively understood in terms of which octahedral rotations or tilts are favored by strain. 

This paper is organized as follows: We start by explaining the crystal structures and important normal modes in Subsection \ref{subsec:rev}. We then present and discuss the strain - tolerance factor phase diagram of HIF RP's in Subsection \ref{subsec:phase}. In Subsection \ref{subsec:switch}, we present the trends of the intrinsic switching barrier as a function of strain. We conclude with a brief summary and discussions in Section \ref{sec:conc}. 

\section{Results}

\subsection{Review of Crystal Structures}
\label{subsec:rev}

\begin{figure}[t]
        \centering
        \includegraphics[width=1.0\linewidth]{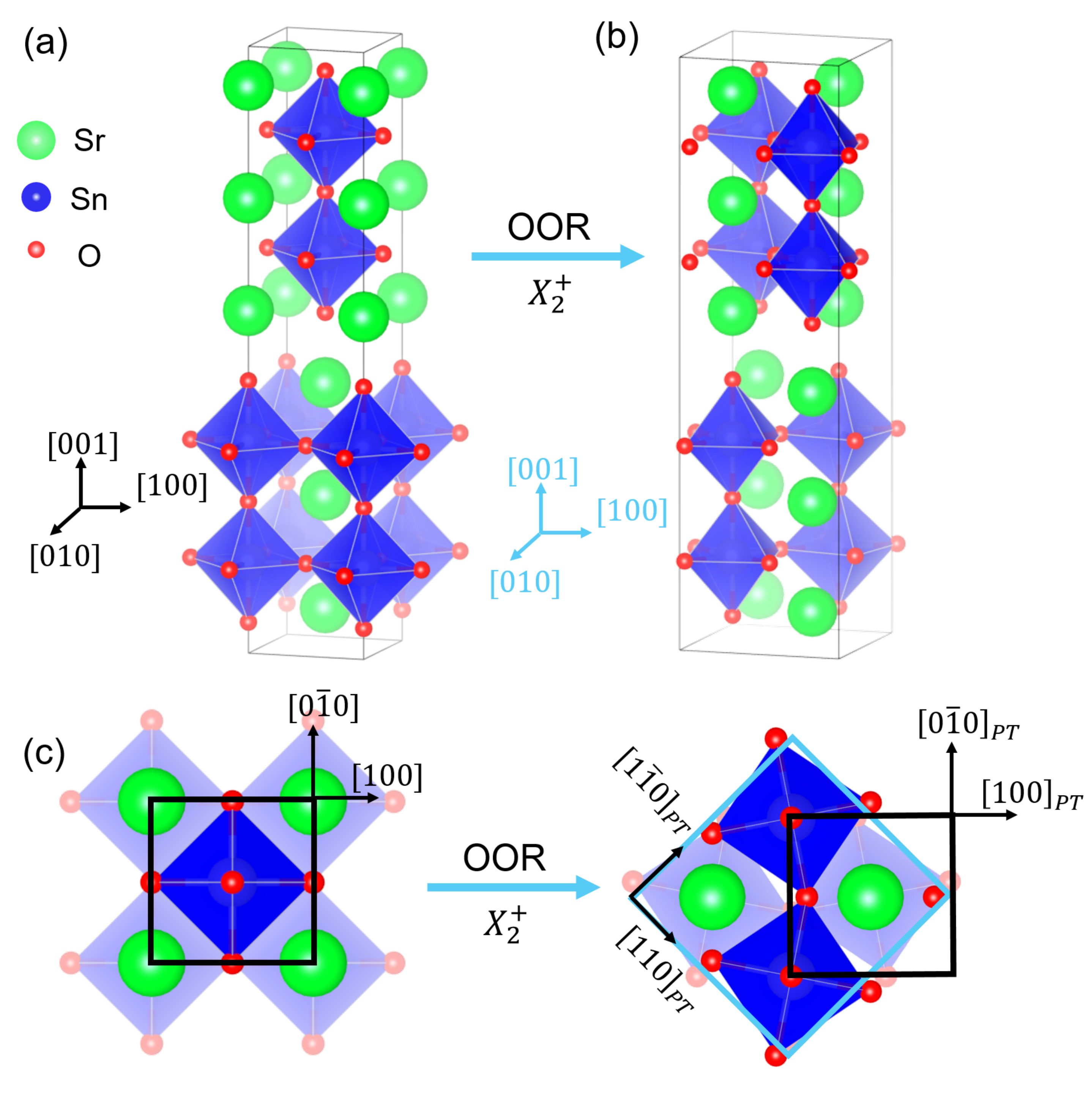}
	\caption{\textbf{The $n=2$ Ruddlesden-Popper Structure} (a) The high symmetry body-centered-tetragonal phase ($I4/mmm$) of A\textsubscript{3}B\textsubscript{2}O\textsubscript{7} RP-phase perovskites. (b) Compounds with tolerance factor less than one develop octahedral rotation/tilt distortions, which are usually associated with normal modes at the X point of the Brillouin zone. (The figure shows the $X_2^+$ mode.) These distortions double the original unit cell and symmetry becomes orthorhombic. (c) Orientations of the crystal axes in the orthorhombic cell are different from those in the high symmetry tetragonal cell. Throughout this paper, we use the axes of a pseudo-tetragonal cell (shown in black) that can be defined within the orthorhombic cell (shown in light blue).}
        \label{fig:structure}
\end{figure}

The A$_3$B$_2$O$_7$ compounds considered in this study are the $n=2$ members of the Ruddlesden-Popper series \cite{Ruddlesden1957,Ruddlesden1958}. They can be considered as layered perovskites with an extra AO layer inserted after every 2 perovskite bi-layers (i.e. 4 atomic layers) along the [001] direction (Fig. \ref{fig:structure}a).  
The extra AO layers cause a shift by ($a/2$, $a/2$, 0) on the $ab$ plane, and hence the structure becomes body centered tetragonal with space group $I4/mmm$ (\#139). This shift also breaks the connectivity of the oxygen octahedra, and the AO double layer is held together by mostly ionic bonds between the A-site cations and O anions. The resulting dimensional reduction has important consequences on the electronic structure and lattice response (For example, Ref.'s \cite{Birol2011, Zhang2013Iridate, Wang2013, Li2019Janotti}). 
	Apart from the dimensional effects, the different periodicity of the Ruddlesden-Popper phases along the layering direction ($c$ axis, or the [001] direction) leads to a smaller Brillouin zone than ABO$_3$ perovskites. The equivalents of various structural instabilities that are at different points of the Brillouin zone in the ABO$_3$ perovskites can fold back onto the same point in A$_3$B$_2$O$_7$ Ruddlesden-Poppers, which leads to interesting couplings between them as discussed below. 
%
%in the next paragraph. Take an example, for ABO$_3$ perovskites with the space group of $Pm\bar{3}m$, phonon modes at the M point and R point will all shift to the X point in the smaller brillouin zone of A$_3$B$_2$O$_7$ Ruddlesden-Poppers, which has a space group of $I4/mmm$. It is not surprising that the M point at (0.5,0.5,0) in the original reciprocal move to the X point at (0.5,0,0) in the new reciprocal space because the a and b-primitive axes get rotated 45 degrees. For the R point at a vertex of cubic Brillouin zone, which is originally at (0.5,0.5,0.5) in the reciprocal space, will move to the X point at an adjacent Brillouin zone because the reduction of dimension in c-axis. 
%
(This point can be qualitatively understood in analogy to a subduction problem, where a zone boundary mode of the parent group corresponds to a zone center mode of the subgroup. For example, when the unitcell of a cubic perovskite is doubled along the [001] axis as a result of cation order, the spacegroup becomes $P4/mmm$ and the zone boundary $X_5^-$ mode splits into $\Gamma_5^-\oplus X_2^-\oplus X_3^-$, where $\Gamma_5^-$ is polar. While there is no direct group-subgroup relationship between the Ruddlesden-Popper and perovskite structures, the $n=2$ Ruddlesden-Poppers have 2 perovskite blocks in their unitcells, and it is thus possible to recognize some phonon modes folded onto the $k_z=0$ plane.)

By far the most common structural distortions that decrease the symmetry of oxide perovskites is the oxygen octahedral rotations: About 90\% of all oxide perovskites have this type of distortion in their crystal structures, which reduces the symmetry of the parent $Pm\bar{3}m$ phase \cite{Lufaso2001}. 
These distortions can be described in terms of symmetry-adapted-modes, which can be classified by irreducible representations (irreps) of the parent spacegroup $Pm\bar{3}m$~\cite{miller1967tables}. The phonon modes that correspond to these distortions are the M point mode M$_2^+$, which is an in-phase rotation of octahedra around one axis, and the R point mode R$_5^-$, which is an out-of-phase rotation of octahedra around one axis. The former is denoted by a `$+$' superscript in the Glazer notation, such as a$^0$a$^0$c$^+$, and the latter is denoted by a `$-$' superscript, such as a$^-$a$^-$a$^-$. The most common rotation pattern that more than half of all oxide perovskites have is a$^-$a$^-$c$^+$, which leads to the space group $Pnma$ (\#62) \cite{Woodward1997}. 
Another distortion that is often significant in the $Pnma$ structure is the X$_5^-$ out-of-phase A-site displacement. Unlike the M$_2^+$ and R$_5^-$, the X$_5^-$ often does not show up as an unstable phonon mode in the high symmetry ($Pm\bar{3}m$) phase. Rather, it is an improper order parameter, which attains a nonzero magnitude only because of a trilinear coupling in the Landau free energy 
\begin{equation}
	\mathcal{F}_{\textrm{trilinear}} = \gamma M_2^+ R_5^- X_5^- .
\end{equation}
The presence of $\mathcal{F}_{\textrm{trilinear}}$ in the free energy expansion, which is imposed by group theory, guarantees a nonzero X$_5^-$ distortion whenever the octahedral rotations M$_2^+$ and R$_5^-$ are present, no matter the sign of the coupling $\gamma$.  

\begin{figure*}
        \centering
        \makebox[\textwidth][c]{
                \includegraphics[width=1.00\linewidth]{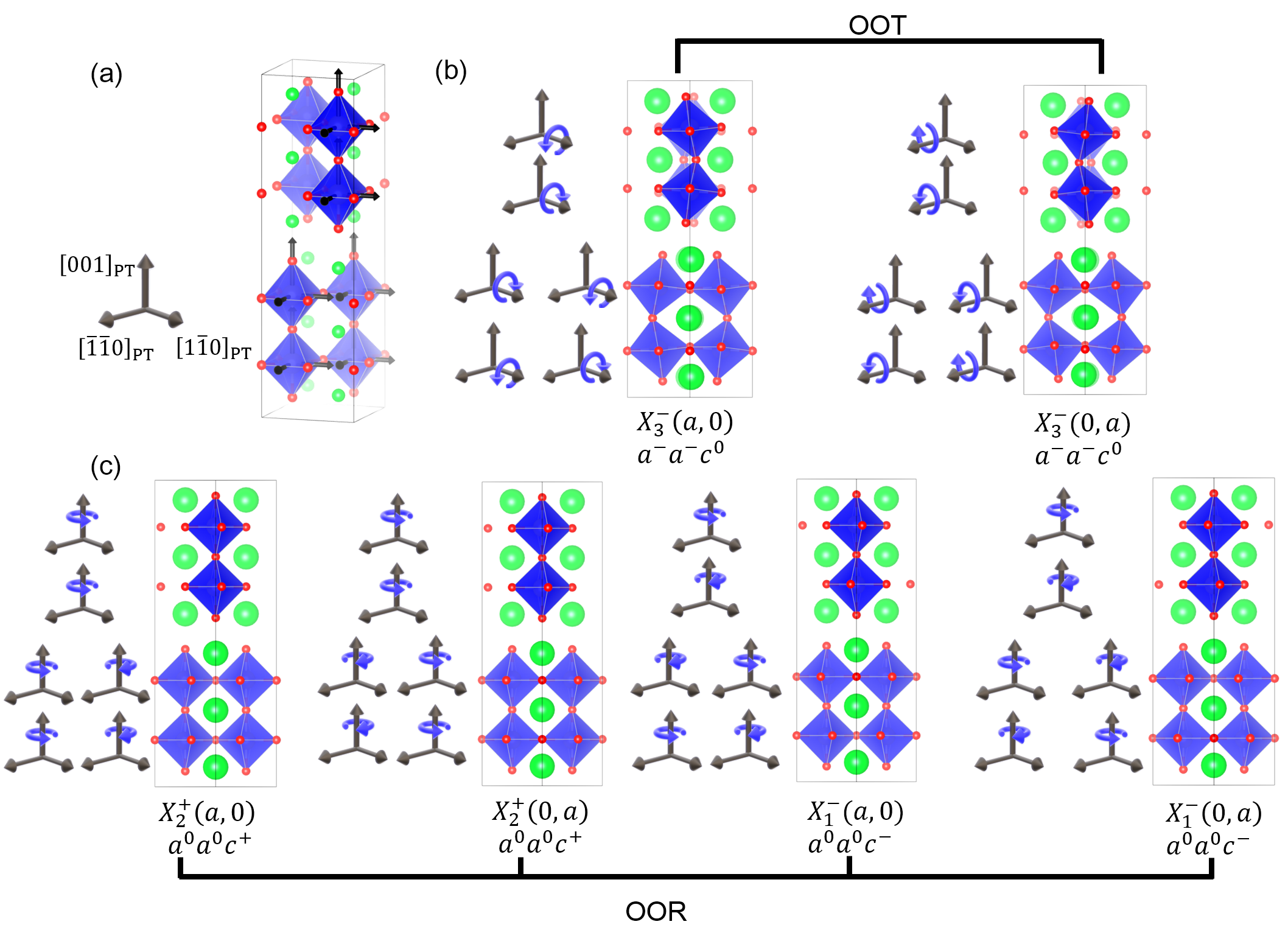}
        }
	\caption{\textbf{The unstable modes at the $X$ point.} (a) The undistorted structure in the orthorhombic supercell. The arrows on the octahedra are along the orthorhombic axes, and are parallel to the arrows in the other panels which denote the direction of octahedral rotations and tilts. (b)-(d) Distortion modes that correspond to different irreps. Both $X_2^+$ and $X_1^-$ modes are rotations around the c-axis. The $X^+_2$ modes are in-phase while $X_1^-$ modes are out-of-phase. The two components of the $X^-_3$ mode are tilts around axes on $ab$ plane. While the words `rotations' and `tilts' are often used interchangeably in the literature, throughout this manuscript we consistently refer to rotations around the c-axis ($X_2^+$ and $X_1^-$) as oxygen octahedral rotations (OOR), and rotations around the axes on the ab plane ($X^-_3$) as oxygen octahedral tilting (OOT). } 
        \label{fig:unstable_modes}
\end{figure*}

Instabilities in the A$_3$B$_2$O$_7$ Ruddlesden-Poppers that are similar to the M$_2^+$ and R$_5^-$ normal modes in the ABO$_3$ perovskites give rise to a wider range of different combinations and resultant symmetries. (For simplicity, we follow the convention to refer octahedral rotations around the out-of-plane ($c$) axis as `rotations' (OOR), and the rotations around the in-plane axes as `tilts' (OOT).) 
One reason for this is that there is a new degree of freedom, since the body-centered primitive cell now contains two oxygen octahedra. Also, the double AO layers break the connectivity of oxygen octahedra, and hence the relative phase of neighboring octahedra on either side of the double layer is not fixed. As an example, we consider the modes relevant to the $A2_1am$ phase observed in Ca$_3$Ti$_2$O$_7$ and many other HIF Ruddlesden-Popper compounds in Fig. \ref{fig:unstable_modes}. 
In ABO$_3$ perovskites, there are two possible rotation patterns around, for example, the $c$ axis: in-phase (M$_2^+$, a$^0$a$^0$c$^+$) or out-of-phase (R$_5^-$, a$^0$a$^0$c$^-$). In the A$_3$B$_2$O$_7$, on the other hand, there are four possibilities: The X$_2^+$ mode corresponds to an in-phase rotation of the two octahedra in one perovskite slab that consists of 5 atomic layers, and is the primitive unit cell. However, X$_2^+$ is a two dimensional irrep, and depending on its direction a particular pair of octahedra on either side of a double AO layer can have either in-phase or out-of-phase rotations, as shown on the left two panels of Fig. \ref{fig:unstable_modes}c. Similarly, the rotations that are out-of-phase within one perovskite slab transform as the two dimensional irrep X$_1^-$, as shown in the right panels of Fig. \ref{fig:unstable_modes}c. 

\begin{figure*}
        \centering
        \includegraphics[width=0.95\linewidth]{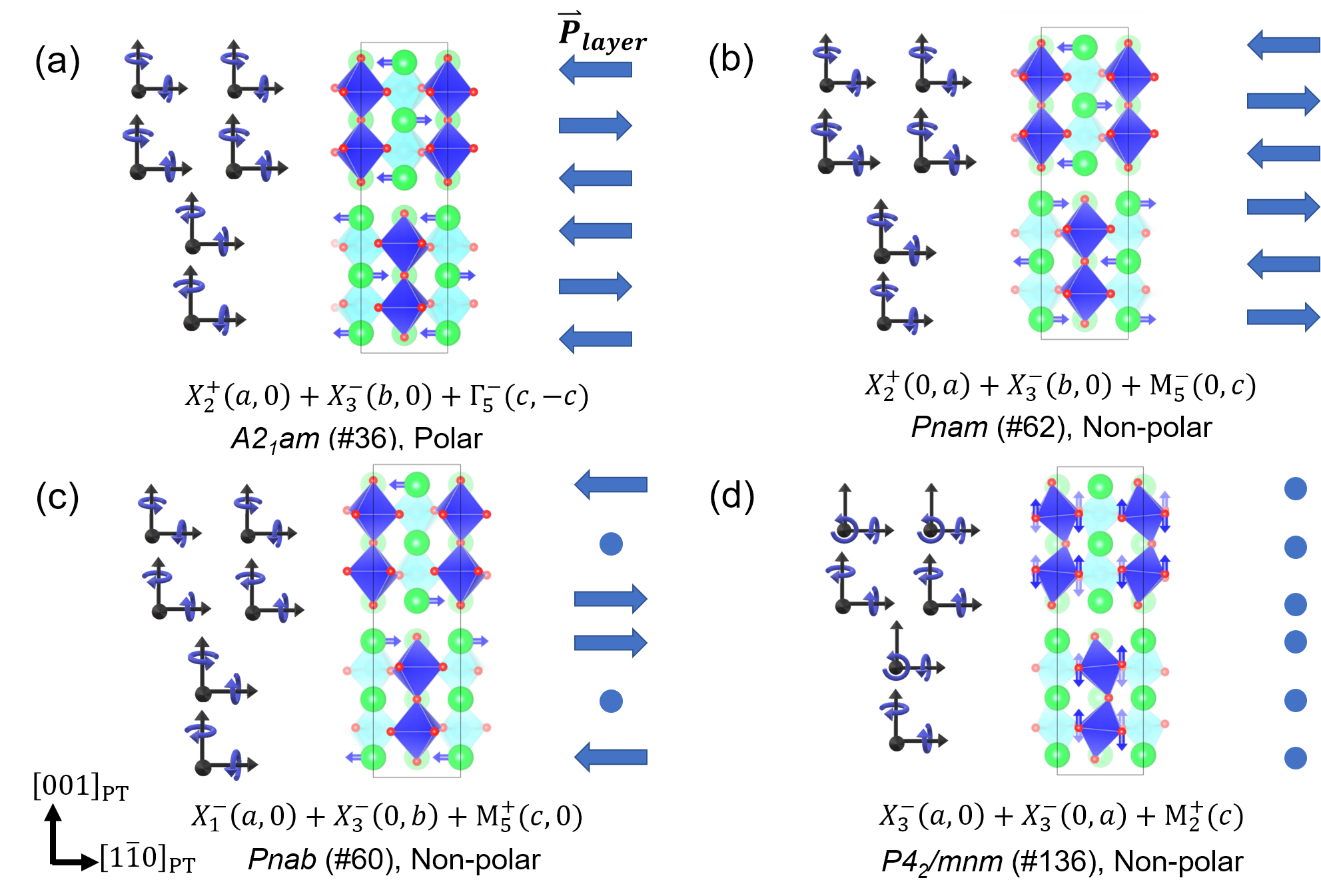}
        \vspace{0.2\baselineskip}
        \caption{\textbf{Possible low energy stable and metastable structures} of RP-phase perovskites A\textsubscript{3}B\textsubscript{2}O\textsubscript{7} with more than one oxygen octahedral rotation modes. Analysis of these modes are presented in Table \ref{tab:trilinear}. }
        \label{fig:ground}
\end{figure*}

The most relevant octahedral rotation modes in  A$_3$B$_2$O$_7$ all have the same wavevector: they correspond to X point normal modes. This leads to a richer set of possibilities for the modes induced by trilinear couplings compared to ABO$_3$ perovskites. In the trilinear coupling terms in ABO$_3$ perovskites, an M and an R mode has to couple with an X mode due to the translational symmetry. 
In A$_3$B$_2$O$_7$ compounds, on the other hand, the trilinear couplings that contain two separate X modes can contain either an M mode or a $\Gamma$ mode as the third mode. (M point is denoted as the Z point in the convention of Ref. \cite{Bradley2010}.) The reason is that there are two separate X points on the Brillouin zone that are related to each other via a four-fold rotation, and depending on which pair of X wavevectors are chosen, their sum can either give the $\Gamma$ or the M wavevector. 
In Table \ref{tab:trilinear}, we list the possible trilinear couplings between two X modes and a third mode in the A$_3$B$_2$O$_7$ structure, and in Fig. \ref{fig:ground}, we display the polarization patterns of some of these structures.

\begin{table*}[h]
	\centering
        \begin{tabular}{|c|c|c|c|}
                \hline
                Irrep 1       & Irrep 2       & Coupled irreps & Space  group                                    \\ 
		\hline
                $X_1^- (a,0)$ &               &               & $Aeaa$ (\#68)                                      \\
                $X_2^+ (a,0)$ &               &               & $Aeam$ (\#64)                                      \\
                $X_3^- (a,0)$ &               &               & $Amam$ (\#63)                                      \\
                $X_3^- (a,a)$ &  & $M_2^+(c)$       & $P4_2/mnm$ (\#136) \\
                $X_1^- (a,0)$ & $X_3^- (0,b)$ & $M_5^+(c,0)$       & $Pnab$ (\#60)                                      \\
                
                $X_1^- (a,0)$ & $X_3^- (b,0)$ & $\Gamma_5^+(c,0) $       & $C/2c$ (\#15)                                      \\
                $X_1^- (a,a)$ & $X_3^- (b,b)$ & $M_1^+(c), M_2^+(c), \Gamma_4^+(c),\Gamma_5^+(c,0)$       & $C/2m$ (\#12) \\
                
                $X_2^+ (a,0)$ & $X_3^- (b,0)$ & $\Gamma_5^-(c,-c)$  & $A2_1am$ (\#36)    \\
                $X_2^+ (0,a)$ & $X_3^- (b,0)$ & $M_5^-(0,c)$       & $Pnam$ (\#62)                                     \\  
                
                $X_2^+ (a,a)$ & $X_3^- (b,b)$ & $M_1^+(c), M_2^+(c) ,\Gamma_4^+(c), \Gamma_5^-(c,0)$       & $C2mm$ (\#38) \\
		\hline
        \end{tabular}
	\caption{List of structures that can be obtained by combining the unstable $X$ modes. The trilinear couplings are obtained using the `Invariants' tool in the Isotropy Software Suite \cite{Hatch2003,ISOTROPY}. While we performed DFT calculations for the energies of all of these phases, only the ones that are close to the lowest energy are shown in the plots. }
    \label{tab:trilinear} 
\end{table*}

Hybrid improper ferroelectricity in the A$_3$B$_2$O$_7$ compounds emerges due to the trilinear coupling between X$_2^+$ and X$_3^-$ modes, which induces a polar displacement $\Gamma_5^-$. In the HIF structure with space group $A2_1am$ (\#36), each AO layer has a polarization, which are in alternating directions within each perovskite slab, and hence cancel each other - but only partially. As a result, every perovskite slab between the double AO layers have a net dipole moment. These moments order in parallel and give rise to a macroscopic polarization (Fig. \ref{fig:ground}a). 
A different combination of the same X modes can couple to the M$_5^-$ mode, leading to anti-parallel slab dipoles, and hence to an anti-polar phase shown in Fig. \ref{fig:ground}b. (We refer to phases with nonzero dipole moments of each perovskite slab as either polar or antipolar.) 
Other combinations of the X modes couple with different M modes, such as M$_5^+$ or M$_2^+$, and give rise to nonpolar phases, where the dipole moments of each atomic layer cancel each other within each perovskite slab between to double AO layers (Fig. \ref{fig:ground}c-d). (We refer to phases where dipole moments of each slab are zero as `nonpolar'.) 
Many of these phases are observed to emerge in various A$_3$B$_2$O$_7$ oxides under biaxial strain or equivalent doping, and are also shown to be important as intermediate states in the coherent switching of polarization \cite{Lu2016, Li2018, Nowadnick2016, Munro2018,Yoshida2018,Li2017}. This is in addition to single-tilt systems observed, for example, at finite temperature \cite{Pitcher2015}. 
In the next subsection, we draw the strain--tolerance factor phase diagram of these compounds to identify regions where these antipolar and nonpolar multi-tilt phases emerge. 

\subsection{Strain Phase Diagram} 
\label{subsec:phase}

\begin{figure}
        \centering
	\includegraphics[width=1.0\linewidth]{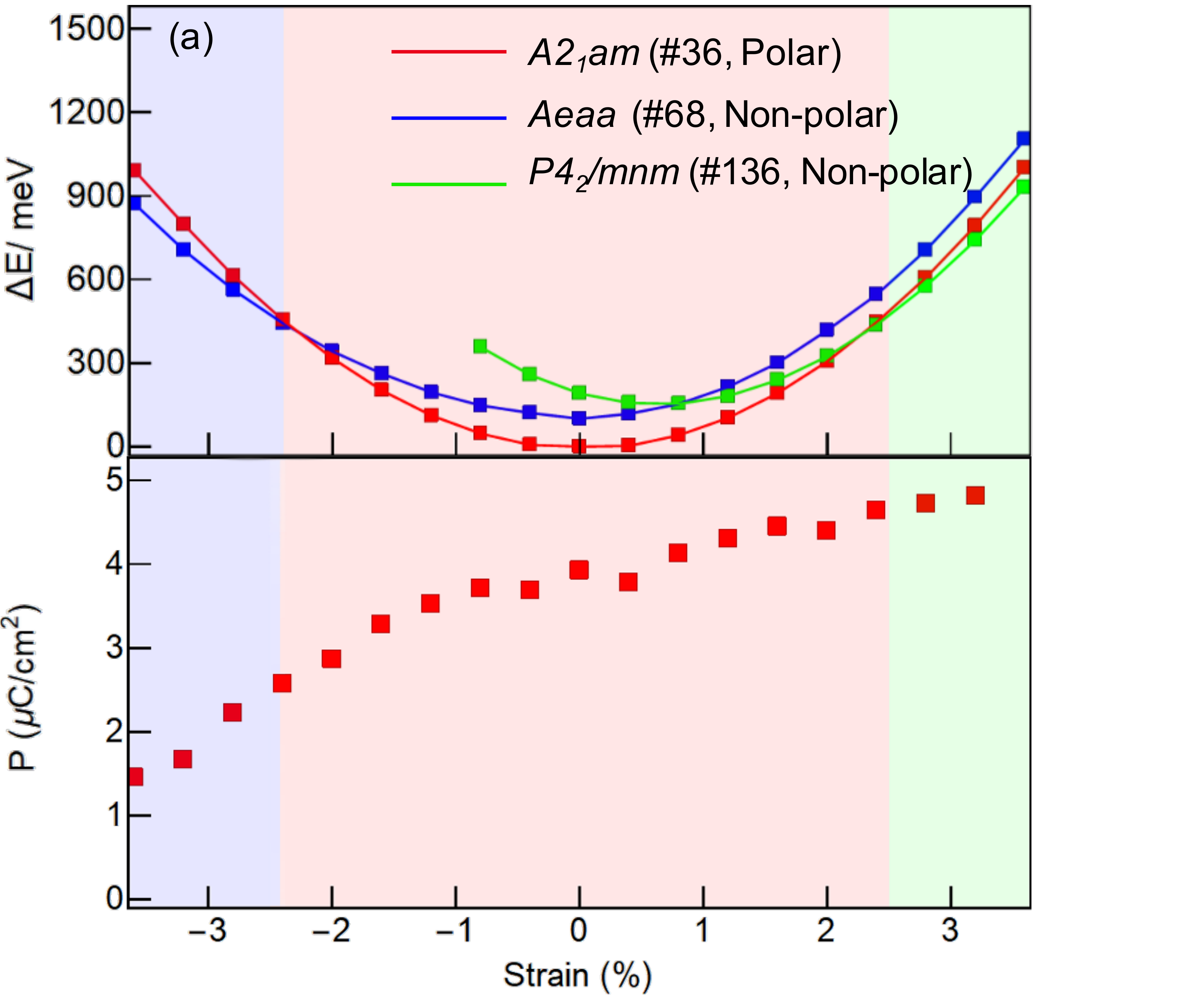}
%        \vspace{0.2\baselineskip}
        \caption{\textbf{Effect of strain on Sr\textsubscript{3}Sn\textsubscript{2}O\textsubscript{7}.}(a) The energy of different metastable phases vary with biaxial strain. Transitions to nonpolar phases are observed on both tensile and compressive strain. (b) The polarization strength of the polar phase as a function of strain. The background colors indicate different ground state structures.}
        \label{fig:SSO_energy}
\end{figure}

Most~--more than half--~of oxide perovskites have a tolerance factor of $\tau<1$, and attain the space group $Pnma$ at low temperatures \cite{Lufaso2001}. The corresponding octahedral rotation pattern a$^-$a$^-$c$^+$ is also common in A$_3$B$_2$O$_7$ Ruddlesden-Poppers, and gives rise to the polar space group $A2_1am$ observed in HIFs. In addition to the polar phase, strain phase diagrams of these compounds often abound with transitions to nonpolar phases introduced in the preceding subsection. 
As an example, in Fig. \ref{fig:SSO_energy}a, we present the energy of three lowest energy structures for Sr$_3$Sn$_2$O$_7$ as a function of biaxial strain \cite{Lu2017}. 
%
%We obtained these energies by relaxing the $c$ lattice constants as well as the internal 
%atomic positions at different values of the in-plane lattice constant $a$. For simplicity, we show energies of 
%only the three relevant phases. 
%
The zero temperature DFT calculations reproduce the experimentally observed room temperature phase $A2_1am$ in the unstrained compound. Both tensile and compressive strain decrease the energy difference between this phase and the next lowest energy state, and there are phase transitions to nonpolar phases for strain $\gtrsim 2.5\%$ on either direction. 
Similar strain driven transitions have been predicted for Sr$_3$Zr$_2$O$_7$ and  Ca$_3$Ti$_2$O$_7$ HIF compounds previously, and the pattern of octahedral rotations often change under strain in the ABO$_3$ compounds as well. A common trend in A$^{2+}$B$^{4+}$O$_3$ perovskites is that tensile biaxial strain suppresses OOR around the out-of-plane axis, whereas compressive strain enhances it. Sr$_3$Sn$_2$O$_7$ follows a similar trend: The transition under tensile strain is to the $P4_2/mnm$ phase, which has only X$_3^-$ tilts, whereas the transition under compressive strain is to the $Aeaa$ phase, which has only the X$_1^-$ rotations around the $c$ axis. 
The transition to these nonpolar phases is not a result of a continuous suppression of polarization by strain: the magnitudes of polarization in the $A2_1am$ phase on both phase boundaries are sizable, and is even enhanced under tensile strain, as shown in Fig. \ref{fig:SSO_energy}(b).

\begin{figure}
\includegraphics[width=1.00\linewidth]{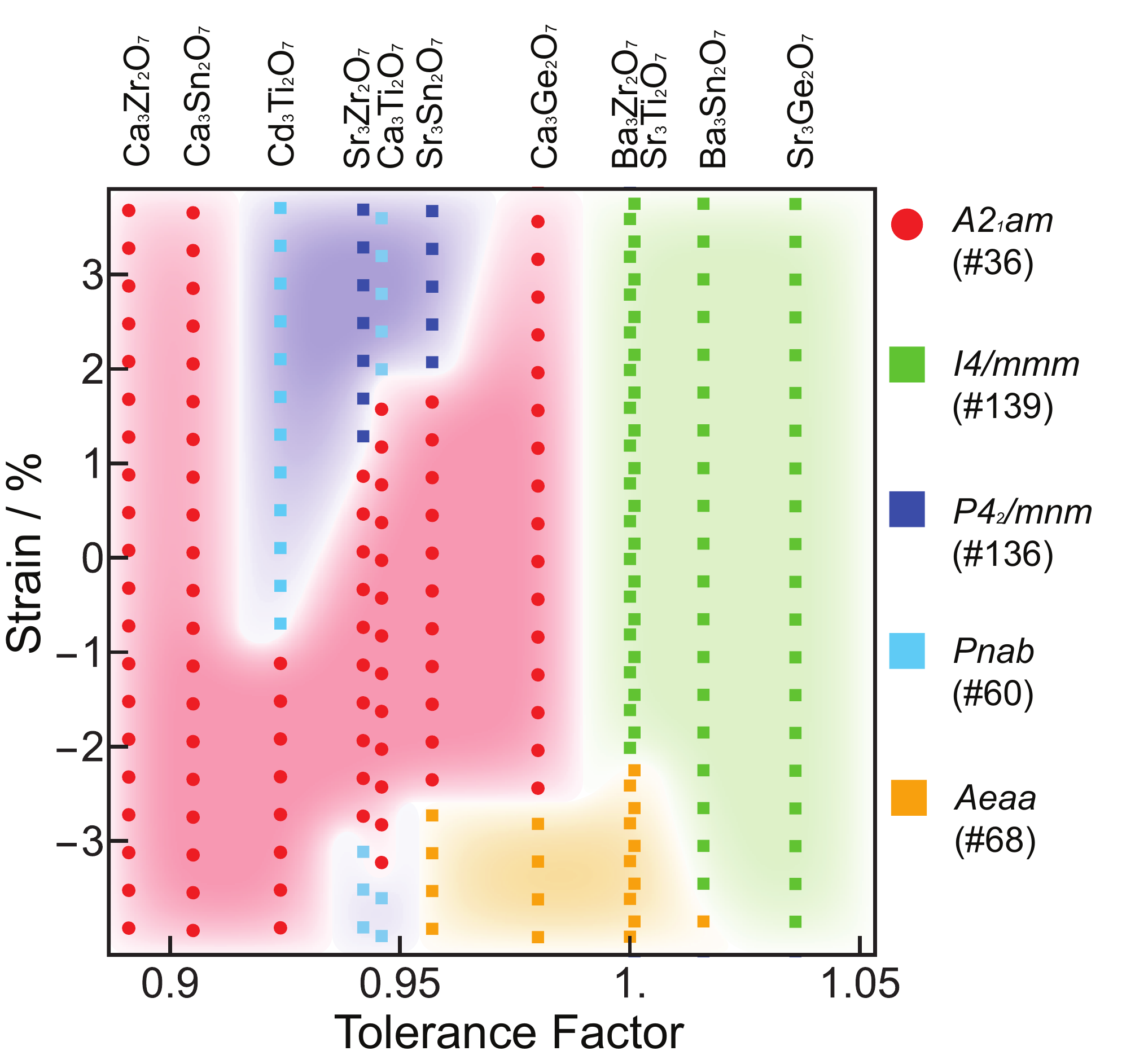}
%\vspace{0.2\baselineskip}
	\centering
	\caption{\textbf{Phase diagram of HIF A\textsubscript{3}B\textsubscript{2}O\textsubscript{7} compounds under biaxial strain.} Red color represents ferroelectric (HIF) phase, the others are all non-polar structures. Results for Ba$_3$Ti$_2$O$_7$ (t=1.06) and Ba$_3$Ge$_2$O$_7$ (t=1.10), which don't display any rotation or tilting, are not shown here. Proper ferroelectric phases of large tolerance factor compounds, such as the one in Sr$_3$Ti$_2$O$_7$ under large tensile strain \cite{Birol2011, Lee2013Birol}, are not displayed either. } 
\label{fig:Phase_diagram}
\end{figure}

\begin{figure*}
              \centering
        \includegraphics[width=5in]{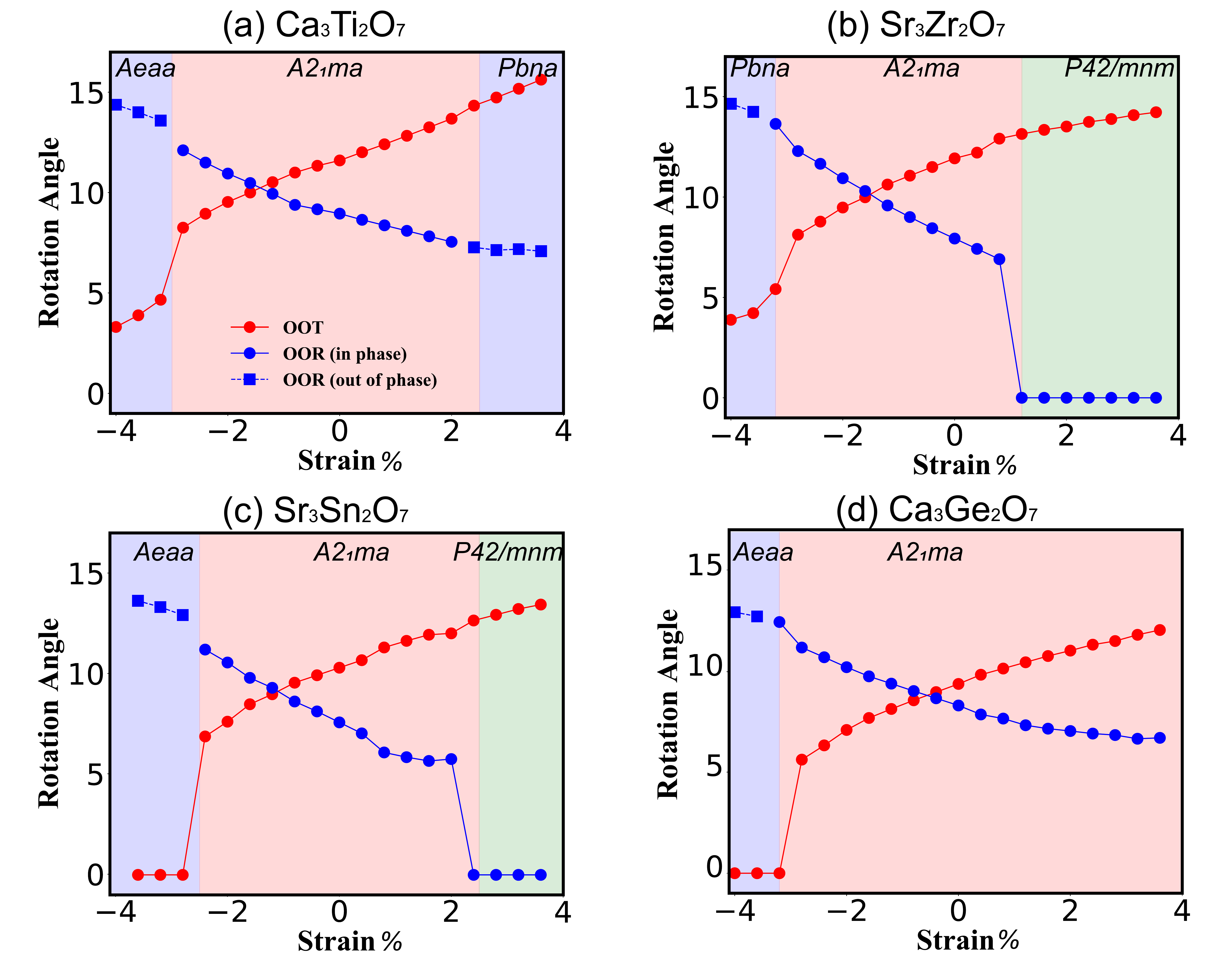}
        \vspace{0.2\baselineskip}
	\caption{\textbf{Effect of Strain on Crystal Structure.} Rotation (OOR) and tilting (OOT) angles as a function of epitaxial strain in (a) Ca\textsubscript{3}Ti\textsubscript{2}O\textsubscript{7}, (b) Sr\textsubscript{3}Zr\textsubscript{2}O\textsubscript{7}, (c) Sr\textsubscript{3}Sn\textsubscript{2}O\textsubscript{7} and (d) Ca\textsubscript{3}Ge\textsubscript{2}O\textsubscript{7}. Different colors represent different phases. Red regions are the ferroelectric phase.}
        \label{fig:angles}
\end{figure*}

In order to elucidate the behavior of different HIF compounds under strain, in Fig. \ref{fig:Phase_diagram} we map out the strain -- tolerance factor phase diagram by considering 11 different A$_3$B$_2$O$_7$ compounds. (We do not include 2 compounds with larger tolerance factors, since they do not display any OOR or OOT. Most of these compounds have been studied from first principles before, but to the best of our knowledge, this is the first time that this information is compiled to display all compounds together.
We consider a strain range of $\pm 4\%$, which covers the experimentally feasible range. For most of the compounds with $\tau<1$ that we consider, the lowest energy unstrained structure is $A2_1am$, which corresponds to the HIF phase. For $0.92\lesssim \tau \lesssim 1$, nonpolar structures emerge under both tensile and compressive strain. We observe three different nonpolar structures: $Pnab$ and $P4_2/mnm$ under tensile strain, and $Pnab$ and $Aeaa$ under compressive strain. They correspond to the following changes in the octahedral rotations and tilts: 
\begin{itemize}
	\item \textbf{Compressive strain induced OOT suppression (leads to $Aeaa$):} This is observed in Sr\textsubscript{3}Sn\textsubscript{2}O\textsubscript{7} and Ca\textsubscript{3}Ge\textsubscript{2}O\textsubscript{7}. The OOT mode amplitude drops to zero and OOR mode phase changes under compressive strain as shown in Fig.~\ref{fig:angles}(c),(d). 
	\item \textbf{Tensile strain induced OOR suppression (leads to $P4_2/mnm$):} This is observed in Sr\textsubscript{3}Zr\textsubscript{2}O\textsubscript{7} and  Sr\textsubscript{3}Sn\textsubscript{2}O\textsubscript{7}. Similar to the first situation, but the OOR mode drops to zero under tensile strain instead of OOT mode, as shown in Fig.~\ref{fig:angles}(b-c).
	\item \textbf{Tensile/compressive strain induced OOR phase change (leads to $Pnab$):} This is observed in Ca\textsubscript{3}Ti\textsubscript{2}O\textsubscript{7} under both tensile and compressive strain, in Sr\textsubscript{3}Zr\textsubscript{2}O\textsubscript{7} under compressive strain, or in Cd\textsubscript{3}Ti\textsubscript{2}O\textsubscript{7} under small tensile as well as compressive strains. (Fig.~\ref{fig:angles}(a-b)). Amplitudes of both the OOR and OOT mode retain non-zero, but the in-phase OOR mode changes into out-of-phase manner. This structure is shown in figure ~\ref{fig:ground}(b). The A-site cations around two interfaces move in the opposite direction, which cancels the polarization in bulk.
\end{itemize}
Some of these transitions are explained by local measures such as the global instability index (GII), which is known to predict the octahedral rotation patterns and angles in ABO$_3$ perovskites successfully \cite{Lufaso2001, Salinas-Sanchez1992}. For example, the transition to $P4_2/mnm$ in Sr$_3$Sn$_2$O$_7$ is coincident with the strain value above which the GII of this phase is the smallest \cite{Supplement}. However, the GII by itself does not explain why the polar $A2_1am$ structure is preferred over the $Aeaa$ one, for these two phases have very similar GII values under compressive strain. It is possible that the interplay of GII with the long-range Coulomb interaction (which is an important factor in stabilizing the polarization in proper ferroelectrics such as BaTiO$_3$\cite{ghosez1996coulomb}) is responsible of the transition to the $Aeaa$ phase. 

The transition to a single-tilt system can be explained phenomenologically by the cross term between OOR and OOT -- a large OOR might suppress OOT and vice versa. All compounds in the $A2_1am$ follow the same aforementioned trend as many ABO$_3$ perovskites that compressive strain enhances OOR, whereas tensile strain enhances OOT (Fig. \ref{fig:angles}a-d). 
(For example, see Ref's \cite{Lu2016, Yang2012, Zayak2006}.) This trend is likely the result of the strain reducing particular B--O bond lengths, which can be increased by the OOT or OOR distortions. 
The lowest order cross term between the OOR and OOT in the free energy is $F\sim \beta R^2 T^2$ (where we denote the amplitudes of rotations and tilts by $R$ and $T$ respectively). For fixed value of $R$, this term renormalizes the coefficient of the $T^2$ term $\sim \alpha T^2$ as $\sim (\alpha+\beta R^2)$, and hence for large OOR $R^2>-\alpha/\beta$, the tilting instability is suppressed, and it becomes energetically favorable to have no tilts, as is the case in compressively strained Sr$_3$Sn$_2$O$_7$ in the $Aeaa$ phase.

A phenomenological explanation of the strain induced transition to nonpolar $Pnab$ structure requires not only the biquadratic terms between the OOR and the OOT modes, but also various trilinear terms that couple these modes to other antiferrodistortive displacements \cite{Lu2016}. It is particularly interesting that in Ca$_3$Ti$_2$O$_7$, this transition is re-entrant in the sense that it happens under both tensile and compressive strains. The GII does not have an obvious trend that explains this transition \cite{Supplement}, and the electrostatic interaction between the O ions on different layers is possibly important \cite{Supplement}. We leave the microscopic explanation of this transition to a future study.

\subsection{Strain tuning of the ferroelectric switching barrier} 
\label{subsec:switch}

Enhanced susceptibilities near second order phase transitions can be exploited to design materials with large responses, for example, magnetic permeability or dielectric constants. While no such enhancement of linear susceptibility is mandated near first order transitions, it is nevertheless possible to obtain large response near a first order phase boundary if the external field can induce the transition. Examples of demonstrations of this approach include Terfenol, Pb(Zr,Ti)O$_3$, and BiFeO$_3$ \cite{Zeches2009, Newnham1998, Birol2012}. The phase boundaries of structural transitions depend on strain very sensitively, and as a result, this approach is a promising means to enhance the response of materials via strain.

\begin{figure*}
	\centering
	\includegraphics[width=0.9\linewidth]{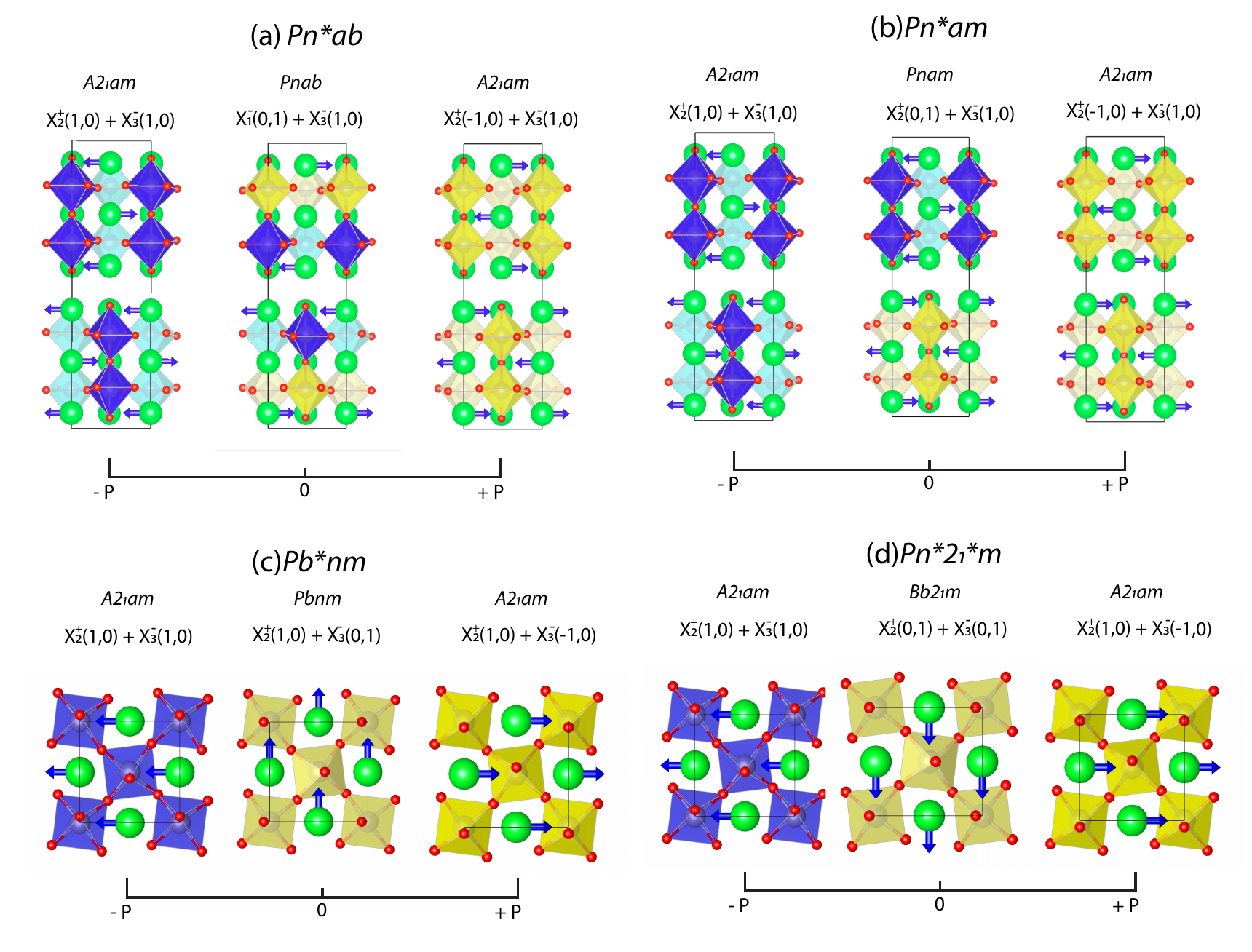}
	\vspace{0.2\baselineskip}
	\caption{\textbf{Four possible polarization switching pathways.} (a)$Pn^*ab$, (b)$Pn^*am$ (c)$Pb^*nm$ (d)$Pn^*2_1^*m$. The octahedra that remain in their original rotation direction are shown in blue, whereas those that switch their rotation direction are shown in yellow.}
	\label{fig:switch_1}
\end{figure*}

\begin{figure}
	\centering
	\includegraphics[width=0.85\linewidth]{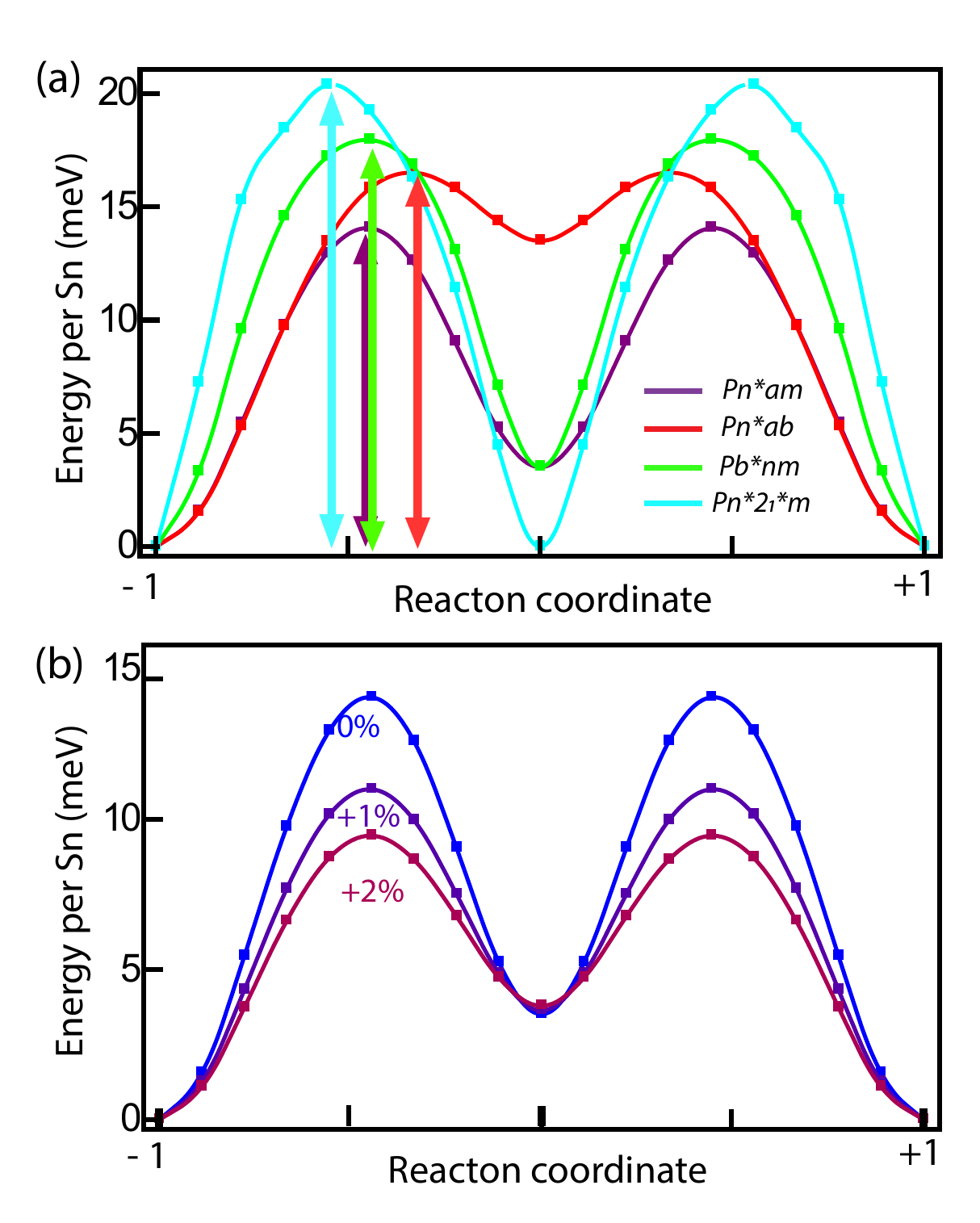}
	\vspace{0.2\baselineskip}
	\caption{\textbf{Energy Barriers for Polarization Switching in Sr$_3$Sn$_2$O$_7$.} (a) The energy barriers of different pathways for unstrained Sr$_3$Sn$_2$O$_7$. The horizontal axis is the ``reaction coordinate'' that parametrizes the switching path. Arrows indicate the barrier heights. (b) The energy of the $Pn^*am$ pathway in Sr$_3$Sn$_2$O$_7$, as a function of tensile biaxial strain.}
	\label{fig:switch_2}
\end{figure}

The question we focus on in this subsection is whether the ferroelectric polarization switching barrier is affected when strain is used to tune the materials to the vicinity of the polar-nonpolar phase transitions.  
In order to answer this question, we use the minimum energy barrier for coherent polarization switching as a proxy to the coercive electric field. While in an actual experiment defects, domain structure, as well as size and shape effects significantly alter the coercive field, trends of coherent switching barrier can be used as a first principles proxy to the trends of the coercive field \cite{Beckman2009} as explicitly shown in HfO$_2$ \cite{Clima2014}. (Finite element methods which take into account the domain structure provide much lower switching barriers \cite{Dittrich2002}.)
In practice, the coherent switching field calculated from the first principles energy barrier by assuming that the dipole moment in every unit cell of an infinite crystal switches at the same time is a gross overestimate. As a result, we don't report the electric field required for switching, but instead report only the energy barriers. 

\begin{figure*}

	\makebox[\textwidth][c]{
		\includegraphics[width=0.95\linewidth]{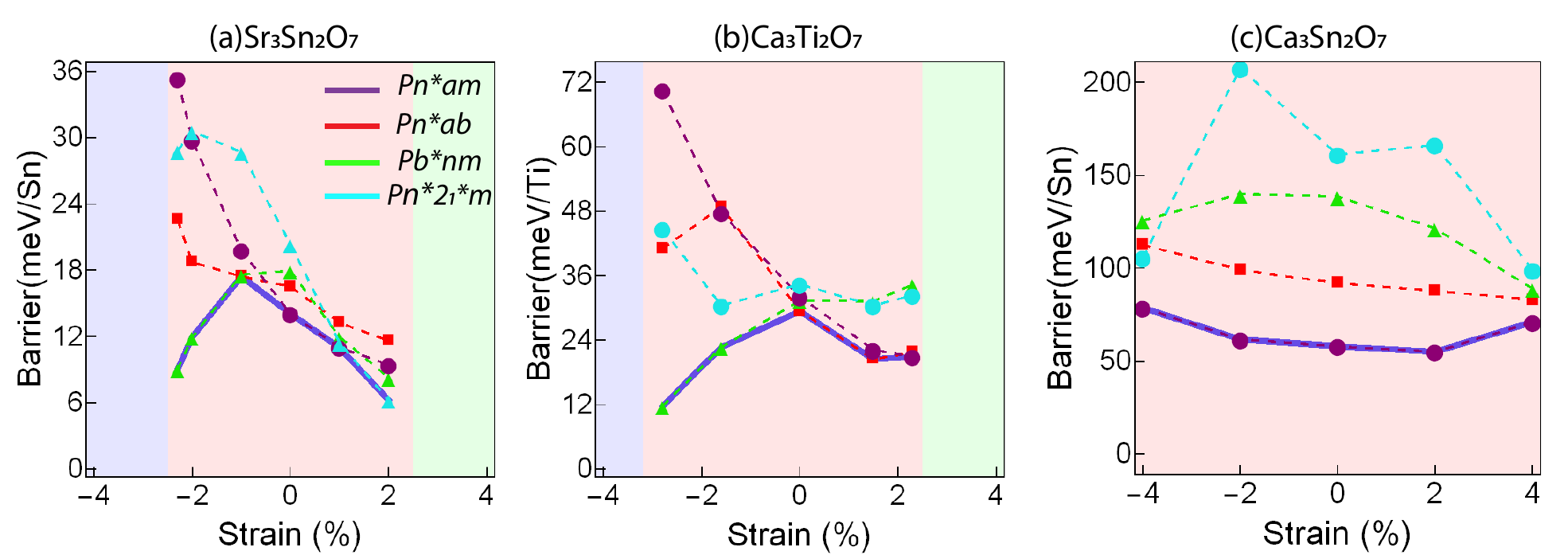}
	}
	%       }
	\vspace{0.2\baselineskip}
	\caption{\textbf{The polarization switching barrier per B-site atom} for (a) Sr\textsubscript{3}Sn\textsubscript{2}O\textsubscript{7}, (b) Ca\textsubscript{3}Ti\textsubscript{2}O\textsubscript{7} and (c) Ca\textsubscript{3}Sn\textsubscript{2}O\textsubscript{7}. The barriers for three distinct pathways are shown here, whilst the thick blue line is the minimum among those three. Background colors indicate different ground states.}
	\label{fig:switch_3}
\end{figure*}

Since in the hybrid improper ferroelectric A\textsubscript{3}B\textsubscript{2}O\textsubscript{7} compounds the polarization emerges as an improper order parameter through a trilinear coupling with rotation and tilting modes, switching one of these two modes is necessary to switch the polarization. It was recognized as early on as in the first HIF paper that this makes different switching pathways possible, and that the corresponding energy barriers can be tuned by strain \cite{Benedek2011}. Later, the work of Nowadnick and Fennie \cite{Nowadnick2016} analyzed the possible roles of different switching mechanisms, and Munro et al. used the idea of distortion symmetry groups to identify other switching pathways \cite{Vanleeuwen2015, Munro2018}. Since then, the energetics of switching in various HIF compounds have been studied, for example in Ref. \cite{Liu2019}. However, to the best of our knowledge, a comparison of different compounds and their strain dependence have not been performed yet. 

In Fig.~\ref{fig:switch_1}(a-d), we show four possible polarization switching pathways.
We follow the convention of the distortion symmetry groups to name these pathways \cite{Vanleeuwen2015}. This process involves identifying not only the symmetry operations shared by all images on the pathway, but also those operations that reverse the distortion, which is the polarization in this case. The latter are referred to as distortion reversal symmetries, and are denoted by a `*' superscript. 
For example, $Pn^*ab$ means that each image along the switching path has two glide planes with translations along a and b axes; and the glide plane $n^*$ reverses the distortion. Three of the switching paths we consider ($Pn^*ab$, $Pb^*nm$ and  $Pn^*am$) have a similar name as their intermediate phase (up to the asterisks), because the spatial symmetry elements of the intermediate phase either remain unchanged or become reversal symmetry operation for other images. But this is not the case for $Pn^*2_1^*m$. 
All of the four are so-called 2-step switching pathways, where there exists a local minimum of energy on the switching path, as seen from Fig.~\ref{fig:switch_2}(a), and they are the lowest ones among such paths for the 3 compounds we considered. They each have distinct intermediate states, but the same initial and final states. 
Since the Ruddlesden-Popper structure consists of weakly bound perovskite blocks separated by an interface between two rock-salt AO layers, it is possible to consider supercells extended along the [001] direction, and polarization being switched in one perovskite block at a time. This, in principle, gives rise to an infinite number of different switching pathways, the barrier energy per formula unit can be arbitrarily small (since only one block out of arbitrarily many switches at each step.) This has been observed in Ref's \cite{Munro2018, Liu2019}, where typically the 4-step switching paths have lower (but comparable) barriers than the 2-step ones, which in turn have lower barriers than the single-step paths. (The path with a very large number of steps can be considered to be a simple model of domain wall in motion along the [001] direction.)
However, this does not necessarily imply that the pathway with the highest number of steps determines the coercive field, because what is more important for the switching under an electric field is the slope of the energy vs. polarization curve \cite{Beckman2009}. For simplicity, as well as computational manageability, we focus only on 2 step switching pathways. 

Each of the four pathways can be reproduced within the same doubled conventional cell as the polar structure. The $Pn^*ab$ and $Pn^*am$ pathways (Fig.~\ref{fig:switch_1}(a-b)) involve changes in the direction of the OOR mode, and both of them have nonpolar intermediate structures, with space groups $Pnab$ and $Pnam$ respectively. The out-of-phase displacements of the A-site cations are along the polar axis in both of these intermediate structures. 
The $Pb^*nm$ pathway involves switching the direction of the OOT mode ($X_3^-$), whereas in the $Pn^*2_1^*m$ both OOR and OOT change directions, as shown in Fig.~\ref{fig:switch_1}(c-d). Mode decompositions of these switching pathways are given in the supplementary information \cite{Supplement}.  

In Fig.~\ref{fig:switch_2}(a), we plot the energy as a function of the reaction coordinate for these four switching pathways in unstrained Sr$_3$Sn$_2$O$_7$. The energy barriers are comparable and the lowest one is for the $Pn^*am$ pathway. Results presented in Fig.~\ref{fig:switch_2}(b) show how the energetics of this path behaves under tensile strain: Tensile strain monotonically decreases the $Pn^*am$ switching barrier, thus lowering the expected coercive field required for switching. 
This is not a surprising result, since the OOR's weaken under tensile strain, as shown in Fig.~\ref{fig:angles}(c) and the $Pn^*am$ pathway involves a change in the OOR character. What is interesting, and important for applications, is that this reduction in the switching barrier is not accompanied with a lower polarization under tensile strain (Fig. \ref{fig:SSO_energy}). Thus, \textit{strain can be used as a means to lower the coercive field of hybrid improper ferroelectrics.}

The strong strain dependence of the switching barrier is not specific only to Sr$_3$Sn$_2$O$_7$, or the $Pn^*am$ pathway. In Fig. \ref{fig:switch_3}(a-b), we show the barrier for different switching paths of Sr$_3$Sn$_2$O$_7$ and Ca$_3$Ti$_2$O$_7$ as a function of strain throughout the strain range that the HIF phase is stable. While the error bars in the energy barriers from the NEB calculations cause the curves to be rather rugged, two trends are evident: \textit{(i)} under tensile strain, the barriers for pathways that involve changing the direction of the OOR mode ($Pn^*am$ and $Pn^*ab$) are lowered, and \textit{(ii)} under compressive strain, the barrier for the pathway that only involve changing the direction of the OOT mode ($Pb^*nm$) is lowered. These are consistent with the tendencies towards OOR and OOT distortions becoming weaker under tensile and compressive strain as discussed earlier. Near 0\% strain, the lowest barrier pathway switches from $Pb^*nm$ or $Pn^*2_1^*m$ to either $Pn^*ab$ or $Pn^*am$, and either strain direction leads to a lower coherent switching energy barrier. The lowest barriers are obtained near the phase boundaries between the polar and nonpolar phases, and the maximum suppression is about 50\% in both compounds.  

Ca$_3$Sn$_2$O$_7$ has a lower tolerance factor than Sr$_3$Sn$_2$O$_7$ and Ca$_3$Ti$_2$O$_7$, and it does not display a strain induced phase transition in the strain range we considered. It does not show a strain induced change in the switching pathway, or a significant decrease in the switching barrier either (Fig.~\ref{fig:switch_3}(c)). This is likely because this compound is very far from the phase boundaries, and with its small tolerance factor, it has such large OOR and OOT that the strain induced changes in the instabilities are inconsequential.

\section{Discussion}
\label{sec:conc}

Since its discovery about a decade ago, hybrid improper ferroelectricity have provided fertile ground for first principles materials by design approaches. Experiments have also been been catching up rapidly, verifying theoretical predictions. 
Multiple hybrid improper ferroelectric Ruddlesden-Popper phases have already been synthesized using bulk methods (for example \cite{Wang2017,Yoshida2018,Yoshida2018-2, Xu2020}). 
Although thin film growth of Ruddlesden-Popper phases, especially for thermodynamically unstable compositions and at large strain values, is usually challenging because of the required stoichiometry control, there has been successful demonstration of switchable HIF in PLD grown films \cite{Li2017}, and both hybrid and conventional oxide molecular beam epitaxy have been used to synthesize phases that are not thermodynamically stable \cite{Lee2013Birol, haislmaier2016creating}. 
Current efforts focus on understanding more than the emergence of ferroelectricity, and to find ways to optimize properties such as the coercive field required for polarization switching. 

In this study, we used first principles calculations to shed light on the strain--tolerance factor phase diagram of $n=2$ Ruddlesden-Popper HIF's, and to come up with a design strategy for obtaining lower coherent switching energy barriers. This quantity, which we used as a proxy for the coercive field, decreases significantly when strain is used to tune the HIF's to the nonpolar phase boundaries, because of the weakening of one of the rotation or tilt modes. We further showed that this weakening, and the resulting decrease in the switching barrier, is not always accompanied with a decrease in the polarization magnitude, for example in Sr$_3$Sn$_2$O$_7$, verifying the point made early on in Ref. \cite{Mulder2013} that a lower barrier does not necessarily mean a lower polarization. 
Our results thus show that biaxial strain, which has historically been used to induce ferroelectricity in many oxides, can also be used as a means to tune the coercive field of hybrid improper ferroelectrics.

\section{Methods}
\subsection{First Principles and Other Calculations} 
Density functional theory calculations are performed using the projector augmented wave approach \cite{blochl1994projector} as implemented in the Vienna Ab-initio Simulation Package (VASP) \cite{VASP1,VASP2}, and using the PBEsol generalized gradient approximation \cite{PBEsol}. All calculations are done in a 48-atom (4 formula unit) supercell, which can be viewed as a $\sqrt{2}\times\sqrt{2}\times{2}$~ multiple of the primitive cell of the reference $I4/mmm$ structure. A $\Gamma$-centered $6\times6\times2$ grid of k-points is used for the Brillouin zone integrals. 

We consider all A$_3$B$_2$O$_7$ compounds with A~=~Ca, Sr, Ba and B~=~Ti, Zr, Sn, Ge, as well as Cd$_3$Ti$_2$O$_7$ \cite{Wang2017,Yoshida2018,Yoshida2018-2,lee2013exploiting,Li2018,Kennedy2011,henriques2007ab,moriwake2011first}. These compounds are all band insulators with sizable gaps, so using the PBEsol generalized gradient approximation is expected to reproduce the crystal structures with reasonable accuracy. 
Biaxial strain boundary conditions are simulated by fixing the in-plane lattice constants, and allowing the out of plane component, as well as internal atomic positions, to relax with an force threshold of 2 $meV$/\AA. The zero strain is defined for each compound by the $a$ lattice constant obtained by completely relaxing the structure in the reference high symmetry structure $I4/mmm$.

The Goldschmidt tolerance factor \cite{Goldschmidt1926}, which is used as a simple measure to predict tendency towards octahedral rotations, and is originally defined in terms of the ionic radii $r$ using  
\begin{equation} 
	\tau = \frac{r_A+r_O}{\sqrt{2} \left(r_B+r_O\right)}
\end{equation}
is instead calculated using the bond lengths for 12 coordinated A-site ($d_{AO}$) and 6 coordinated B-site ($d_{BO}$) ions from the bond valence model as 
\begin{equation}
	\tau=\frac{d_{AO}}{\sqrt{2}\cdot d_{BO}} .
\end{equation}
(This approach is following Ref. \cite{Lufaso2001}.)

In order to calculate the minimum energy barrier for polarization switching, climbing-image nudged elastic band (CI-NEB) method was used to further relax linearly interpolated switching paths to the minimum energy path \cite{NEB}. The spring constant was set to $5$~eV/\AA$^2$, and a convergence criterion of $1$~meV per supercell was used. 
Distortion symmetry groups \cite{Vanleeuwen2015, Padmanabhan2020} are used to enumerate and name the possible initial pathways following Ref. \cite{Munro2018} with the help of the DiSPy package \cite{Munro2019}. 
All the switching pathways reported in the text retain their symmetry for all values of the reaction coordinate under NEB calculation. 

As various points in this paper, symmetry and group theory related arguments are built using the Isotropy Software Package\cite{isotropy2007} and the Bilbao Crystallographic Server\cite{bilbao1,bilbao2,bilbao3}. VESTA software was used for visualization of crystal structures.\cite{VESTA2008}

\section{Data Availability} 
Data for the phase diagram and the switching paths are available at the Data Repository for University of Minnesota at https://doi.org/10.13020/hvr3-bg02 .

\section{Acknowledgements}
This work was supported primarily by the National Science Foundation through the University of Minnesota MRSEC under Award Number DMR-2011401. We acknowledge the Minnesota Supercomputing Institute (MSI) at the University of Minnesota for providing resources that contributed to the research results reported within this paper.

\section{Author Contributions}
Both authors contributed to the planning of the project, and writing of the manuscript. First principles calculations were performed by S.L.

\section{Competing Interests}
The authors declare no competing financial or non-financial interests.

\bibliographystyle{naturemag}

\begin{thebibliography}{10}
\expandafter\ifx\csname url\endcsname\relax
  \def\url#1{\texttt{#1}}\fi
\expandafter\ifx\csname urlprefix\endcsname\relax\def\urlprefix{URL }\fi
\providecommand{\bibinfo}[2]{#2}
\providecommand{\eprint}[2][]{\url{#2}}

\bibitem{Benedek2013}
\bibinfo{author}{Benedek, N.~A.} \& \bibinfo{author}{Fennie, C.~J.}
\newblock \bibinfo{title}{{Why are there so few perovskite ferroelectrics?}}
\newblock \emph{\bibinfo{journal}{Journal of Physical Chemistry C}}
  \textbf{\bibinfo{volume}{117}}, \bibinfo{pages}{13339--13349}
  (\bibinfo{year}{2013}).

\bibitem{Benedek2011}
\bibinfo{author}{Benedek, N.~A.} \& \bibinfo{author}{Fennie, C.~J.}
\newblock \bibinfo{title}{{Hybrid improper ferroelectricity: A mechanism for
  controllable polarization-magnetization coupling}}.
\newblock \emph{\bibinfo{journal}{Physical Review Letters}}
  \textbf{\bibinfo{volume}{106}}, \bibinfo{pages}{107204}
  (\bibinfo{year}{2011}).

\bibitem{Oh2015}
\bibinfo{author}{Oh, Y.~S.}, \bibinfo{author}{Luo, X.}, \bibinfo{author}{Huang,
  F.-T.}, \bibinfo{author}{Wang, Y.} \& \bibinfo{author}{Cheong, S.-W.}
\newblock \bibinfo{title}{{Experimental demonstration of hybrid improper
  ferroelectricity and the presence of abundant charged walls in (Ca,Sr)3Ti2O7
  crystals }}.
\newblock \emph{\bibinfo{journal}{Nature materials}}
  \textbf{\bibinfo{volume}{14}}, \bibinfo{pages}{407--413}
  (\bibinfo{year}{2015}).

\bibitem{Wang2017}
\bibinfo{author}{Wang, Y.}, \bibinfo{author}{Huang, F.~T.},
  \bibinfo{author}{Luo, X.}, \bibinfo{author}{Gao, B.} \&
  \bibinfo{author}{Cheong, S.~W.}
\newblock \bibinfo{title}{{The First Room-Temperature Ferroelectric Sn
  Insulator and Its Polarization Switching Kinetics}}.
\newblock \emph{\bibinfo{journal}{Advanced Materials}}
  \textbf{\bibinfo{volume}{29}} (\bibinfo{year}{2017}).

\bibitem{Yoshida2018}
\bibinfo{author}{Yoshida, S.} \emph{et~al.}
\newblock \bibinfo{title}{{Hybrid Improper Ferroelectricity in (Sr,Ca)3Sn2O7
  and Beyond: Universal Relationship between Ferroelectric Transition
  Temperature and Tolerance Factor in n=2 Ruddlesden–Popper Phases}}.
\newblock \emph{\bibinfo{journal}{Journal of the American Chemical Society}}
  \textbf{\bibinfo{volume}{140}}, \bibinfo{pages}{15690--15700}
  (\bibinfo{year}{2018}).

\bibitem{Yoshida2018-2}
\bibinfo{author}{Yoshida, S.} \emph{et~al.}
\newblock \bibinfo{title}{{Ferroelectric Sr3Zr2O7: Competition between Hybrid
  Improper Ferroelectric and Antiferroelectric Mechanisms}}.
\newblock \emph{\bibinfo{journal}{Advanced Functional Materials}}
  \textbf{\bibinfo{volume}{28}}, \bibinfo{pages}{1801856}
  (\bibinfo{year}{2018}).

\bibitem{Pitcher2015}
\bibinfo{author}{Pitcher, M.~J.} \emph{et~al.}
\newblock \bibinfo{title}{{Tilt engineering of spontaneous polarization and
  magnetization above 300K in a bulk layered perovskite}}.
\newblock \emph{\bibinfo{journal}{Science}} \textbf{\bibinfo{volume}{347}},
  \bibinfo{pages}{420--424} (\bibinfo{year}{2015}).

\bibitem{Elcombe1991}
\bibinfo{author}{Elcombe, M.~M.} \emph{et~al.}
\newblock \bibinfo{title}{{Structure determinations for Ca3Ti2O7, Ca4Ti3O10,
  Ca3.6Sr0.4Ti3O10 and a refinement of Sr3Ti2O7}}.
\newblock \emph{\bibinfo{journal}{Acta Crystallographica Section B}}
  \textbf{\bibinfo{volume}{47}}, \bibinfo{pages}{305--314}
  (\bibinfo{year}{1991}).

\bibitem{Mulder2013}
\bibinfo{author}{Mulder, A.~T.}, \bibinfo{author}{Benedek, N.~A.},
  \bibinfo{author}{Rondinelli, J.~M.} \& \bibinfo{author}{Fennie, C.~J.}
\newblock \bibinfo{title}{{Turning ABO3 antiferroelectrics into ferroelectrics:
  Design rules for practical rotation-driven ferroelectricity in double
  perovskites and A3B2O7 Ruddlesden-popper compounds}}.
\newblock \emph{\bibinfo{journal}{Advanced Functional Materials}}
  \textbf{\bibinfo{volume}{23}}, \bibinfo{pages}{4810--4820}
  (\bibinfo{year}{2013}).

\bibitem{Lines2001Book}
\bibinfo{author}{Lines, M.} \& \bibinfo{author}{Glass, A.}
\newblock \emph{\bibinfo{title}{Principles and Applications of Ferroelectrics
  and Related Materials}}.
\newblock International series of monographs on physics
  (\bibinfo{publisher}{OUP Oxford}, \bibinfo{year}{2001}).

\bibitem{Xu2020}
\bibinfo{author}{Xu, X.} \emph{et~al.}
\newblock \bibinfo{title}{{Highly Tunable Ferroelectricity in Hybrid Improper
  Ferroelectric Sr3Sn2O7}}.
\newblock \emph{\bibinfo{journal}{Advanced Functional Materials}}
  \textbf{\bibinfo{volume}{2003623}}, \bibinfo{pages}{1--9}
  (\bibinfo{year}{2020}).

\bibitem{Scott2007}
\bibinfo{author}{Scott, J.~F.}
\newblock \bibinfo{title}{{Applications of modern ferroelectrics}}.
\newblock \emph{\bibinfo{journal}{Science}} \textbf{\bibinfo{volume}{315}},
  \bibinfo{pages}{954--959} (\bibinfo{year}{2007}).

\bibitem{Cole2014}
\bibinfo{author}{Cole, J.}, \bibinfo{author}{Ahmed, S.~J.},
  \bibinfo{author}{Curiel, L.}, \bibinfo{author}{Pichardo, S.} \&
  \bibinfo{author}{Rubel, O.}
\newblock \bibinfo{title}{{Marble game with optimal ferroelectric switching}}.
\newblock \emph{\bibinfo{journal}{Journal of Physics: Condensed Matter}}
  \textbf{\bibinfo{volume}{26}}, \bibinfo{pages}{135901}
  (\bibinfo{year}{2014}).

\bibitem{Xu2018}
\bibinfo{author}{Xu, R.} \emph{et~al.}
\newblock \bibinfo{title}{{Reducing Coercive-Field Scaling in Ferroelectric
  Thin Films via Orientation Control}}.
\newblock \emph{\bibinfo{journal}{ACS Nano}} \textbf{\bibinfo{volume}{12}},
  \bibinfo{pages}{4736--4743} (\bibinfo{year}{2018}).

\bibitem{Liu2018}
\bibinfo{author}{Liu, Z.~Q.} \emph{et~al.}
\newblock \bibinfo{title}{{Electrically reversible cracks in an intermetallic
  film controlled by an electric field}}.
\newblock \emph{\bibinfo{journal}{Nature Communications}}
  \textbf{\bibinfo{volume}{9}}, \bibinfo{pages}{1--7} (\bibinfo{year}{2018}).

\bibitem{Li2017}
\bibinfo{author}{Li, X.} \emph{et~al.}
\newblock \bibinfo{title}{{Ultra-low coercive field of improper ferroelectric
  Ca3Ti2O7 epitaxial thin films}}.
\newblock \emph{\bibinfo{journal}{Applied Physics Letters}}
  \textbf{\bibinfo{volume}{110}}, \bibinfo{pages}{42901}
  (\bibinfo{year}{2017}).

\bibitem{Wang2018}
\bibinfo{author}{Wang, T.} \emph{et~al.}
\newblock \bibinfo{title}{{Engineering SrSnO 3 Phases and Electron Mobility at
  Room Temperature Using Epitaxial Strain}}.
\newblock \emph{\bibinfo{journal}{ACS Applied Materials and Interfaces}}
  \textbf{\bibinfo{volume}{10}}, \bibinfo{pages}{43802--43808}
  (\bibinfo{year}{2018}).

\bibitem{Chaturvedi2020}
\bibinfo{author}{Chaturvedi, V.} \emph{et~al.}
\newblock \bibinfo{title}{Strain-induced majority carrier inversion in
  ferromagnetic epitaxial
  $\mathrm{LaCo}{\mathrm{o}}_{3\ensuremath{-}\ensuremath{\delta}}$ thin films}.
\newblock \emph{\bibinfo{journal}{Phys. Rev. Materials}}
  \textbf{\bibinfo{volume}{4}}, \bibinfo{pages}{034403} (\bibinfo{year}{2020}).

\bibitem{Clima2014}
\bibinfo{author}{Clima, S.} \emph{et~al.}
\newblock \bibinfo{title}{{Identification of the ferroelectric switching
  process and dopant-dependent switching properties in orthorhombic HfO2: A
  first principles insight}}.
\newblock \emph{\bibinfo{journal}{Applied Physics Letters}}
  \textbf{\bibinfo{volume}{104}} (\bibinfo{year}{2014}).

\bibitem{Lu2016}
\bibinfo{author}{Lu, X.~Z.} \& \bibinfo{author}{Rondinelli, J.~M.}
\newblock \bibinfo{title}{{Epitaxial-strain-induced polar-to-nonpolar
  transitions in layered oxides}}.
\newblock \emph{\bibinfo{journal}{Nature Materials}}
  \textbf{\bibinfo{volume}{15}}, \bibinfo{pages}{951--955}
  (\bibinfo{year}{2016}).

\bibitem{Nowadnick2016}
\bibinfo{author}{Nowadnick, E.~A.} \& \bibinfo{author}{Fennie, C.~J.}
\newblock \bibinfo{title}{{Domains and ferroelectric switching pathways in
  Ca3Ti2O7 from first principles}}.
\newblock \emph{\bibinfo{journal}{Physical Review B}}
  \textbf{\bibinfo{volume}{94}}, \bibinfo{pages}{104105}
  (\bibinfo{year}{2016}).

\bibitem{Munro2018}
\bibinfo{author}{Munro, J.~M.} \emph{et~al.}
\newblock \bibinfo{title}{{Discovering minimum energy pathways via distortion
  symmetry groups}}.
\newblock \emph{\bibinfo{journal}{Physical Review B}}
  \textbf{\bibinfo{volume}{98}}, \bibinfo{pages}{85107} (\bibinfo{year}{2018}).

\bibitem{Ruddlesden1957}
\bibinfo{author}{Ruddlesden, S.~N.} \& \bibinfo{author}{Popper, P.}
\newblock \bibinfo{title}{{New compounds of the K2NiF4 type}}.
\newblock \emph{\bibinfo{journal}{Acta Crystallographica}}
  \textbf{\bibinfo{volume}{10}}, \bibinfo{pages}{538--539}
  (\bibinfo{year}{1957}).

\bibitem{Ruddlesden1958}
\bibinfo{author}{Ruddlesden, S.~N.} \& \bibinfo{author}{Popper, P.}
\newblock \bibinfo{title}{{The compound Sr3Ti2O7 and its structure}}.
\newblock \emph{\bibinfo{journal}{Acta Crystallographica}}
  \textbf{\bibinfo{volume}{11}}, \bibinfo{pages}{54--55}
  (\bibinfo{year}{1958}).

\bibitem{Birol2011}
\bibinfo{author}{Birol, T.}, \bibinfo{author}{Benedek, N.~A.} \&
  \bibinfo{author}{Fennie, C.~J.}
\newblock \bibinfo{title}{{Interface Control of Emergent Ferroic Order in
  Ruddlesden-Popper Srn+1TinO3n+1}}.
\newblock \emph{\bibinfo{journal}{Physical Review Letters}}
  \textbf{\bibinfo{volume}{107}}, \bibinfo{pages}{257602}
  (\bibinfo{year}{2011}).
\newblock \eprint{1110.1122}.

\bibitem{Zhang2013Iridate}
\bibinfo{author}{Zhang, H.}, \bibinfo{author}{Haule, K.} \&
  \bibinfo{author}{Vanderbilt, D.}
\newblock \bibinfo{title}{{Effective J=1/2 Insulating State in
  Ruddlesden-Popper Iridates: An LDA+DMFT Study}}.
\newblock \emph{\bibinfo{journal}{Physical Review Letters}}
  \textbf{\bibinfo{volume}{111}}, \bibinfo{pages}{246402}
  (\bibinfo{year}{2013}).

\bibitem{Wang2013}
\bibinfo{author}{Wang, Q.} \emph{et~al.}
\newblock \bibinfo{title}{Dimensionality-controlled mott transition and
  correlation effects in single-layer and bilayer perovskite iridates}.
\newblock \emph{\bibinfo{journal}{Phys. Rev. B}} \textbf{\bibinfo{volume}{87}},
  \bibinfo{pages}{245109} (\bibinfo{year}{2013}).

\bibitem{Li2019Janotti}
\bibinfo{author}{Li, W.} \emph{et~al.}
\newblock \bibinfo{title}{{Band gap evolution in Ruddlesden-Popper phases}}.
\newblock \emph{\bibinfo{journal}{Physical Review Materials}}
  \textbf{\bibinfo{volume}{3}}, \bibinfo{pages}{101601} (\bibinfo{year}{2019}).
\newblock \eprint{1905.02598}.

\bibitem{Lufaso2001}
\bibinfo{author}{Lufaso, M.~W.} \& \bibinfo{author}{Woodward, P.~M.}
\newblock \bibinfo{title}{{Prediction of the crystal structures of perovskites
  using the software program SPuDS}}.
\newblock \emph{\bibinfo{journal}{Acta Crystallographica Section B: Structural
  Science}} \textbf{\bibinfo{volume}{57}}, \bibinfo{pages}{725--738}
  (\bibinfo{year}{2001}).

\bibitem{miller1967tables}
\bibinfo{author}{Miller, S.~C.} \& \bibinfo{author}{Love, W.~F.}
\newblock \emph{\bibinfo{title}{Tables of irreducible representations of space
  groups and co-representations of magnetic space groups}}
  (\bibinfo{publisher}{Pruett Press}, \bibinfo{year}{1967}).

\bibitem{Woodward1997}
\bibinfo{author}{Woodward, P.~M.}
\newblock \bibinfo{title}{{Octahedral Tilting in Perovskites. II. Structure
  Stabilizing Forces}}.
\newblock \emph{\bibinfo{journal}{Acta Crystallographica Section B: Structural
  Science}} \textbf{\bibinfo{volume}{53}}, \bibinfo{pages}{44--66}
  (\bibinfo{year}{1997}).

\bibitem{Bradley2010}
\bibinfo{author}{Bradley, C.} \& \bibinfo{author}{Cracknell, A.}
\newblock \emph{\bibinfo{title}{The Mathematical Theory of Symmetry in Solids:
  Representation Theory for Point Groups and Space Groups}}.
\newblock EBSCO ebook academic collection (\bibinfo{publisher}{OUP Oxford},
  \bibinfo{year}{2010}).

\bibitem{Hatch2003}
\bibinfo{author}{Hatch, D.~M.} \& \bibinfo{author}{Stokes, H.~T.}
\newblock \bibinfo{title}{{INVARIANTS : program for obtaining a list of
  invariant polynomials of the order-parameter components associated with
  irreducible representations of a space group }}.
\newblock \emph{\bibinfo{journal}{Journal of Applied Crystallography}}
  \textbf{\bibinfo{volume}{36}}, \bibinfo{pages}{951--952}
  (\bibinfo{year}{2003}).

\bibitem{ISOTROPY}
\bibinfo{author}{Stokes, H.~T.}, \bibinfo{author}{Hatch, D.~M.} \&
  \bibinfo{author}{Campbell, B.}
\newblock \emph{\bibinfo{title}{ISOTROPY Software Suite, iso.byu.edu.}}

\bibitem{Li2018}
\bibinfo{author}{Li, C.~F.} \emph{et~al.}
\newblock \bibinfo{title}{{Structural transitions in hybrid improper
  ferroelectric Ca3Ti2O7 tuned by site-selective isovalent substitutions: A
  first-principles study}}.
\newblock \emph{\bibinfo{journal}{Physical Review B}}
  \textbf{\bibinfo{volume}{97}} (\bibinfo{year}{2018}).

\bibitem{Lu2017}
\bibinfo{author}{Lu, X.~Z.} \& \bibinfo{author}{Rondinelli, J.~M.}
\newblock \bibinfo{title}{{Room Temperature Electric-Field Control of Magnetism
  in Layered Oxides with Cation Order}}.
\newblock \emph{\bibinfo{journal}{Advanced Functional Materials}}
  \textbf{\bibinfo{volume}{27}}, \bibinfo{pages}{1604312}
  (\bibinfo{year}{2017}).

\bibitem{Salinas-Sanchez1992}
\bibinfo{author}{Salinas-Sanchez, A.}, \bibinfo{author}{Garcia-Mu{\~{n}}oz,
  J.~L.}, \bibinfo{author}{Rodriguez-Carvajal, J.},
  \bibinfo{author}{Saez-Puche, R.} \& \bibinfo{author}{Martinez, J.~L.}
\newblock \bibinfo{title}{{Structural characterization of R2BaCuO5 (R = Y, Lu,
  Yb, Tm, Er, Ho, Dy, Gd, Eu and Sm) oxides by X-ray and neutron diffraction}}.
\newblock \emph{\bibinfo{journal}{Journal of Solid State Chemistry}}
  \textbf{\bibinfo{volume}{100}}, \bibinfo{pages}{201--211}
  (\bibinfo{year}{1992}).

\bibitem{Supplement}
\emph{\bibinfo{title}{See the supplemental information for details.}}

\bibitem{ghosez1996coulomb}
\bibinfo{author}{Ghosez, P.}, \bibinfo{author}{Gonze, X.} \&
  \bibinfo{author}{Michenaud, J.-P.}
\newblock \bibinfo{title}{{Coulomb interaction and ferroelectric instability of
  BaTiO3}}.
\newblock \emph{\bibinfo{journal}{EPL (Europhysics Letters)}}
  \textbf{\bibinfo{volume}{33}}, \bibinfo{pages}{713} (\bibinfo{year}{1996}).

\bibitem{Lee2013Birol}
\bibinfo{author}{Lee, C.-H.~H.} \emph{et~al.}
\newblock \bibinfo{title}{{Exploiting dimensionality and defect mitigation to
  create tunable microwave dielectrics}}.
\newblock \emph{\bibinfo{journal}{Nature}} \textbf{\bibinfo{volume}{502}},
  \bibinfo{pages}{532--536} (\bibinfo{year}{2013}).

\bibitem{Yang2012}
\bibinfo{author}{Yang, Y.}, \bibinfo{author}{Ren, W.}, \bibinfo{author}{Wang,
  D.} \& \bibinfo{author}{Bellaiche, L.}
\newblock \bibinfo{title}{{Understanding and Revisiting Properties of EuTiO3
  Bulk Material and Films from First Principles}}.
\newblock \emph{\bibinfo{journal}{Physical Review Letters}}
  \textbf{\bibinfo{volume}{109}}, \bibinfo{pages}{267602}
  (\bibinfo{year}{2012}).

\bibitem{Zayak2006}
\bibinfo{author}{Zayak, A.~T.}, \bibinfo{author}{Huang, X.},
  \bibinfo{author}{Neaton, J.~B.} \& \bibinfo{author}{Rabe, K.~M.}
\newblock \bibinfo{title}{{Structural, electronic, and magnetic properties of
  SrRuO3 under epitaxial strain}}.
\newblock \emph{\bibinfo{journal}{Physical Review B}}
  \textbf{\bibinfo{volume}{74}}, \bibinfo{pages}{094104}
  (\bibinfo{year}{2006}).

\bibitem{Zeches2009}
\bibinfo{author}{Zeches, R.~J.} \emph{et~al.}
\newblock \bibinfo{title}{{A strain-driven morphotropic phase boundary in
  BiFeO3}}.
\newblock \emph{\bibinfo{journal}{Science}} \textbf{\bibinfo{volume}{326}},
  \bibinfo{pages}{977--980} (\bibinfo{year}{2009}).

\bibitem{Newnham1998}
\bibinfo{author}{Newnham, R.~E.}
\newblock \bibinfo{title}{{Phase Transformations in Smart Materials}}.
\newblock \emph{\bibinfo{journal}{Acta Crystallographica Section A Foundations
  of Crystallography}} \textbf{\bibinfo{volume}{54}}, \bibinfo{pages}{729--737}
  (\bibinfo{year}{1998}).

\bibitem{Birol2012}
\bibinfo{author}{Birol, T.} \emph{et~al.}
\newblock \bibinfo{title}{{The magnetoelectric effect in transition metal
  oxides: Insights and the rational design of new materials from first
  principles}}.
\newblock \emph{\bibinfo{journal}{Current Opinion in Solid State and Materials
  Science}} \textbf{\bibinfo{volume}{16}}, \bibinfo{pages}{227--242}
  (\bibinfo{year}{2012}).
\newblock \eprint{1207.5026}.

\bibitem{Beckman2009}
\bibinfo{author}{Beckman, S.~P.}, \bibinfo{author}{Wang, X.},
  \bibinfo{author}{Rabe, K.~M.} \& \bibinfo{author}{Vanderbilt, D.}
\newblock \bibinfo{title}{{Ideal barriers to polarization reversal and
  domain-wall motion in strained ferroelectric thin films}}.
\newblock \emph{\bibinfo{journal}{Physical Review B - Condensed Matter and
  Materials Physics}} \textbf{\bibinfo{volume}{79}} (\bibinfo{year}{2009}).

\bibitem{Dittrich2002}
\bibinfo{author}{Dittrich, R.} \emph{et~al.}
\newblock \bibinfo{title}{{A path method for finding energy barriers and
  minimum energy paths in complex micromagnetic systems}}.
\newblock \emph{\bibinfo{journal}{Journal of Magnetism and Magnetic Materials}}
  \textbf{\bibinfo{volume}{250}}, \bibinfo{pages}{12--19}
  (\bibinfo{year}{2002}).

\bibitem{Vanleeuwen2015}
\bibinfo{author}{Vanleeuwen, B.~K.} \& \bibinfo{author}{Gopalan, V.}
\newblock \bibinfo{title}{{The antisymmetry of distortions}}.
\newblock \emph{\bibinfo{journal}{Nature Communications}}
  \textbf{\bibinfo{volume}{6}}, \bibinfo{pages}{8818} (\bibinfo{year}{2015}).

\bibitem{Liu2019}
\bibinfo{author}{Liu, X.~Q.}, \bibinfo{author}{Lu, J.~J.},
  \bibinfo{author}{Chen, B.~H.}, \bibinfo{author}{Zhang, B.~H.} \&
  \bibinfo{author}{Chen, X.~M.}
\newblock \bibinfo{title}{{Hybrid improper ferroelectricity and possible
  ferroelectric switching paths in Sr3Hf2O7}}.
\newblock \emph{\bibinfo{journal}{Journal of Applied Physics}}
  \textbf{\bibinfo{volume}{125}}, \bibinfo{pages}{114105}
  (\bibinfo{year}{2019}).

\bibitem{haislmaier2016creating}
\bibinfo{author}{Haislmaier, R.~C.}, \bibinfo{author}{Stone, G.},
  \bibinfo{author}{Alem, N.} \& \bibinfo{author}{Engel-Herbert, R.}
\newblock \bibinfo{title}{Creating ruddlesden-popper phases by hybrid molecular
  beam epitaxy}.
\newblock \emph{\bibinfo{journal}{Applied Physics Letters}}
  \textbf{\bibinfo{volume}{109}}, \bibinfo{pages}{043102}
  (\bibinfo{year}{2016}).

\bibitem{blochl1994projector}
\bibinfo{author}{Bl{\"o}chl, P.~E.}
\newblock \bibinfo{title}{Projector augmented-wave method}.
\newblock \emph{\bibinfo{journal}{Physical review B}}
  \textbf{\bibinfo{volume}{50}}, \bibinfo{pages}{17953} (\bibinfo{year}{1994}).

\bibitem{VASP1}
\bibinfo{author}{Kresse, G.} \& \bibinfo{author}{Joubert, D.}
\newblock \bibinfo{title}{From ultrasoft pseudopotentials to the projector
  augmented-wave method}.
\newblock \emph{\bibinfo{journal}{Phys. Rev. B}} \textbf{\bibinfo{volume}{59}},
  \bibinfo{pages}{1758--1775} (\bibinfo{year}{1999}).

\bibitem{VASP2}
\bibinfo{author}{Kresse, G.} \& \bibinfo{author}{Hafner, J.}
\newblock \bibinfo{title}{Ab initio molecular dynamics for liquid metals}.
\newblock \emph{\bibinfo{journal}{Phys. Rev. B}} \textbf{\bibinfo{volume}{47}},
  \bibinfo{pages}{558--561} (\bibinfo{year}{1993}).

\bibitem{PBEsol}
\bibinfo{author}{Perdew, J.~P.} \emph{et~al.}
\newblock \bibinfo{title}{Restoring the density-gradient expansion for exchange
  in solids and surfaces}.
\newblock \emph{\bibinfo{journal}{Phys. Rev. Lett.}}
  \textbf{\bibinfo{volume}{100}}, \bibinfo{pages}{136406}
  (\bibinfo{year}{2008}).

\bibitem{lee2013exploiting}
\bibinfo{author}{Lee, C.-H.} \emph{et~al.}
\newblock \bibinfo{title}{Exploiting dimensionality and defect mitigation to
  create tunable microwave dielectrics}.
\newblock \emph{\bibinfo{journal}{Nature}} \textbf{\bibinfo{volume}{502}},
  \bibinfo{pages}{532} (\bibinfo{year}{2013}).

\bibitem{Kennedy2011}
\bibinfo{author}{Kennedy, B.~J.}, \bibinfo{author}{Zhou, Q.} \&
  \bibinfo{author}{Avdeev, M.}
\newblock \bibinfo{title}{{The ferroelectric phase of CdTiO3: A powder neutron
  diffraction study}}.
\newblock \emph{\bibinfo{journal}{Journal of Solid State Chemistry}}
  \textbf{\bibinfo{volume}{184}}, \bibinfo{pages}{2987--2993}
  (\bibinfo{year}{2011}).

\bibitem{henriques2007ab}
\bibinfo{author}{Henriques, J.}, \bibinfo{author}{Caetano, E.},
  \bibinfo{author}{Freire, V.}, \bibinfo{author}{da~Costa, J.} \&
  \bibinfo{author}{Albuquerque, E.}
\newblock \bibinfo{title}{{Ab initio structural, electronic and optical
  properties of orthorhombic CaGeO3}}.
\newblock \emph{\bibinfo{journal}{Journal of Solid State Chemistry}}
  \textbf{\bibinfo{volume}{180}}, \bibinfo{pages}{974--980}
  (\bibinfo{year}{2007}).

\bibitem{moriwake2011first}
\bibinfo{author}{Moriwake, H.} \emph{et~al.}
\newblock \bibinfo{title}{{First-principles calculations of lattice dynamics in
  CdTiO 3 and CaTiO 3: Phase stability and ferroelectricity}}.
\newblock \emph{\bibinfo{journal}{Physical Review B}}
  \textbf{\bibinfo{volume}{84}}, \bibinfo{pages}{104114}
  (\bibinfo{year}{2011}).

\bibitem{Goldschmidt1926}
\bibinfo{author}{Goldschmidt, V.~M.}
\newblock \bibinfo{title}{Die gesetze der krystallochemie}.
\newblock \emph{\bibinfo{journal}{Naturwissenschaften}}
  \textbf{\bibinfo{volume}{14}}, \bibinfo{pages}{477--485}
  (\bibinfo{year}{1926}).

\bibitem{NEB}
\bibinfo{author}{Henkelman, G.} \& \bibinfo{author}{J{\'o}nsson, H.}
\newblock \bibinfo{title}{Improved tangent estimate in the nudged elastic band
  method for finding minimum energy paths and saddle points}.
\newblock \emph{\bibinfo{journal}{The Journal of chemical physics}}
  \textbf{\bibinfo{volume}{113}}, \bibinfo{pages}{9978--9985}
  (\bibinfo{year}{2000}).

\bibitem{Padmanabhan2020}
\bibinfo{author}{Padmanabhan, H.}, \bibinfo{author}{Munro, J.~M.},
  \bibinfo{author}{Dabo, I.} \& \bibinfo{author}{Gopalan, V.}
\newblock \bibinfo{title}{{Antisymmetry: Fundamentals and Applications}}.
\newblock \emph{\bibinfo{journal}{Annual Review of Materials Research}}
  \textbf{\bibinfo{volume}{50}},
  \bibinfo{pages}{annurev--matsci--100219--101404} (\bibinfo{year}{2020}).

\bibitem{Munro2019}
\bibinfo{author}{Munro, J.~M.}, \bibinfo{author}{Liu, V.~S.},
  \bibinfo{author}{Gopalan, V.} \& \bibinfo{author}{Dabo, I.}
\newblock \bibinfo{title}{{Implementation of distortion symmetry for the nudged
  elastic band method with DiSPy}}.
\newblock \emph{\bibinfo{journal}{npj Computational Materials}}
  \textbf{\bibinfo{volume}{5}}, \bibinfo{pages}{52} (\bibinfo{year}{2019}).

\bibitem{isotropy2007}
\bibinfo{author}{Stokes, H.}, \bibinfo{author}{Hatch, D.} \&
  \bibinfo{author}{Campbell, B.}
\newblock \bibinfo{title}{Isotropy} (\bibinfo{year}{2007}).

\bibitem{bilbao1}
\bibinfo{author}{Aroyo, M.~I.}, \bibinfo{author}{Kirov, A.},
  \bibinfo{author}{Capillas, C.}, \bibinfo{author}{Perez-Mato, J.~M.} \&
  \bibinfo{author}{Wondratschek, H.}
\newblock \bibinfo{title}{Bilbao crystallographic server. ii. representations
  of crystallographic point groups and space groups}.
\newblock \emph{\bibinfo{journal}{Acta Crystallographica Section A}}
  \textbf{\bibinfo{volume}{62}}, \bibinfo{pages}{115--128}
  (\bibinfo{year}{2006}).

\bibitem{bilbao2}
\bibinfo{author}{Aroyo, M.} \emph{et~al.}
\newblock \bibinfo{title}{Bilbao crystallographic server: I. databases and
  crystallographic computing programs}.
\newblock \emph{\bibinfo{journal}{Zeitschrift fur Kristallographie}}
  \textbf{\bibinfo{volume}{221}}, \bibinfo{pages}{15--27}
  (\bibinfo{year}{2006}).

\bibitem{bilbao3}
\bibinfo{author}{Aroyo, M.~I.} \emph{et~al.}
\newblock \bibinfo{title}{Crystallography online: Bilbao crystallographic
  server}.
\newblock \emph{\bibinfo{journal}{Bulgarian Chemical Communications}}
  \textbf{\bibinfo{volume}{43}}, \bibinfo{pages}{183--197}
  (\bibinfo{year}{2011}).

\bibitem{VESTA2008}
\bibinfo{author}{Momma, K.} \& \bibinfo{author}{Izumi, F.}
\newblock \emph{\bibinfo{journal}{Journal of Applied Crystallography}}
  \textbf{\bibinfo{volume}{41}}, \bibinfo{pages}{653--658}
  (\bibinfo{year}{2008}).

\end{thebibliography}

\end{document}